\newcommand*{\TechReport}{}%
\newcommand*{\AcmFormat}{}%

\ifdefined\TechReport
\documentclass[10pt,sigconf,letterpaper]{acmart}
\else
\documentclass[10pt,sigconf,letterpaper,anonymous]{acmart}
\fi

\usepackage{times}  
\ifdefined\AcmFormat
\else
\usepackage[hidelinks]{hyperref}

\usepackage{graphicx}
\usepackage{cite}
\usepackage[cmex10]{amsmath}
\usepackage{amssymb}
\usepackage{comment}
\usepackage{float}
\usepackage{subcaption}
\usepackage{url}
\usepackage{amsfonts}
\usepackage{amsthm}

\fi
\usepackage{balance}

\ifdefined\AcmFormat
\else
\fi

\usepackage{array}
\usepackage{subcaption}
\usepackage{enumitem}

\newlength{\grafflecm}
\ifdefined\AcmFormat

\renewcommand\footnotetextcopyrightpermission[1]{} 
\fi
\begin{document}

\makeatletter
\ifdefined\AcmFormat
\renewcommand\@formatdoi[1]{\ignorespaces}
\makeatother
\fi
\pagestyle{plain}

\date{}

\title{
Using Internet Measurements to Map \\ the 2022 Ukrainian Refugee Crisis
}

\ifdefined\AcmFormat

\author{Tal Mizrahi}
\affiliation{%
  \institution{Technion --- Israel Institute of Technology}
}

\author{Jose Yallouz}
\affiliation{%
  \institution{Technion --- Israel Institute of Technology}
}

\else

\fi

\thispagestyle{empty}

\ifdefined\AcmFormat
\begin{abstract}
The conflict in Ukraine, starting in February 2022, began the largest refugee crisis in decades, with millions of Ukrainian refugees crossing the border to neighboring countries and millions of others forced to move within the country. In this paper we present an insight into how Internet measurements can be used to analyze the refugee crisis. Based on preliminary data from the first two months of the war we analyze how measurement data indicates the trends in the flow of refugees from Ukraine to its neighboring countries, and onward to other countries. We believe that these insights can greatly contribute to the ongoing international effort to map the flow of refugees in order to aid and protect them. 
\end{abstract}

\maketitle

\section{Introduction}
\label{IntroSec}
The 2022 war in Ukraine, which started with Russia's invasion on 24 February 2022, caused Europe's largest refugee crisis since World War II~\cite{UkrainianExodus}.
The unprecedented refugee crisis has been an increasing concern, with nearly 6 million refugees exiting Ukraine as of May 2022~\cite{UNHCR} and over a quarter of the Ukrainian population being displaced~\cite{Quarter} inside the country. The United Nations High Commissioner for Refugees (UNHCR) is chartered to aid and protect refugees in this crisis. 

The UNHCR publishes data~\cite{UNHCR} about refugees that exited Ukraine to one of its seven neighboring countries on a daily basis.
The accumulated number of refugees since the beginning of the war is illustrated in Figure~\ref{fig:Ref}, and a histogram of the neighboring countries to which the refugees crossed the border is shown in Figure~\ref{fig:RefHist}. 

One of the first steps in aiding these refugees is being able to monitor the distribution of refugees throughout the world. However, while the UNHCR continuously keeps track of the influx from Ukraine to the neighboring countries, once exiting Ukraine individuals may move freely between countries and therefore it is difficult to have a clear picture of the refugee distribution throughout the world, and especially throughout Europe since border crossing between EU countries is not monitored.

\begin{figure}[!t]
  \centering
  \begin{subfigure}[t]{.24\textwidth}
  \centering
  \fbox{\includegraphics[trim={4cm 1cm 6cm 1cm},clip,height=6\grafflecm]{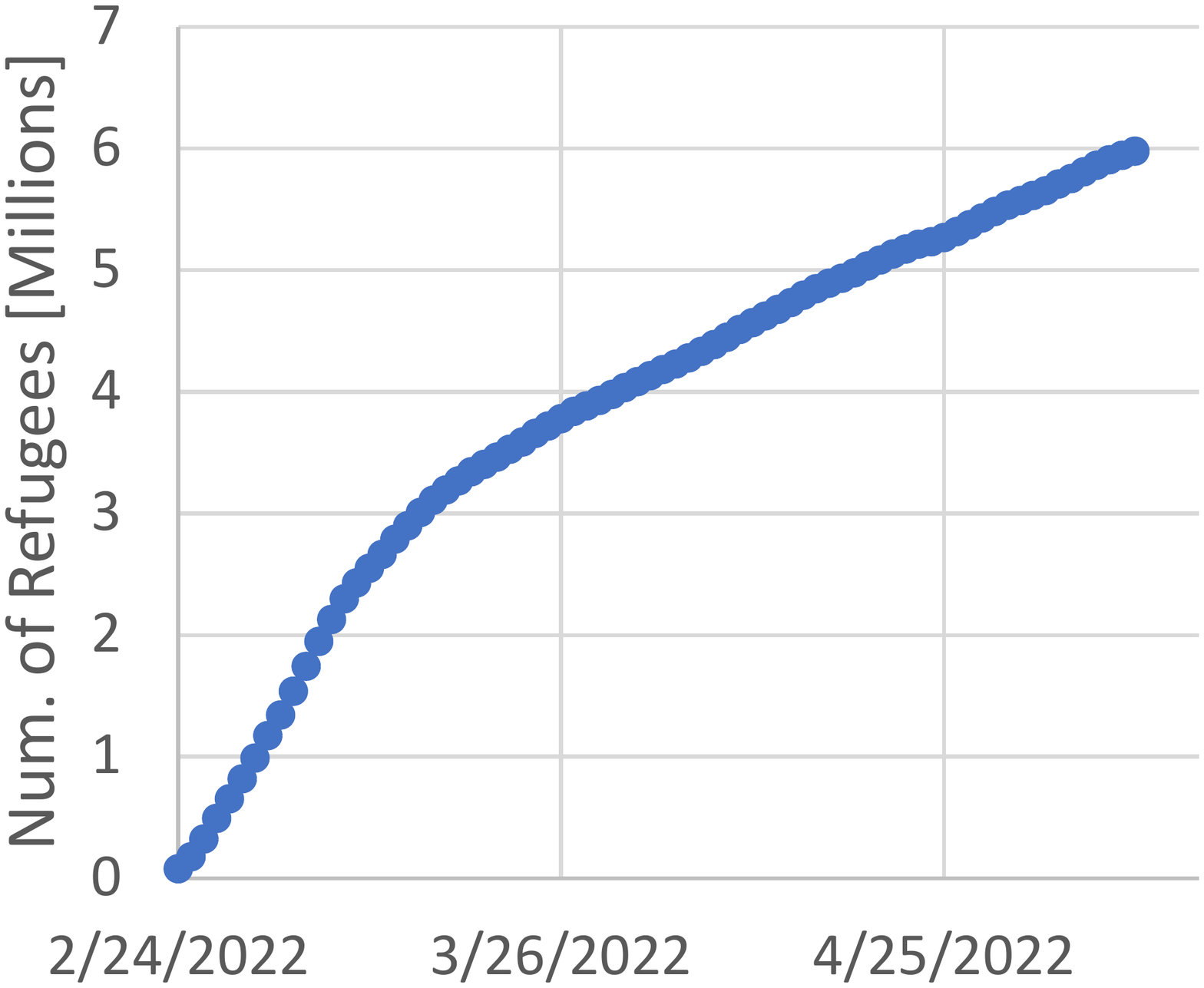}}
	\captionsetup{justification=centering}
  \caption{Total number of refugees}
  \label{fig:Ref}
  \end{subfigure}%
  \begin{subfigure}[t]{.24\textwidth}
  \centering
  \fbox{\includegraphics[trim={6cm 1cm 5cm 1cm},clip,height=6\grafflecm]{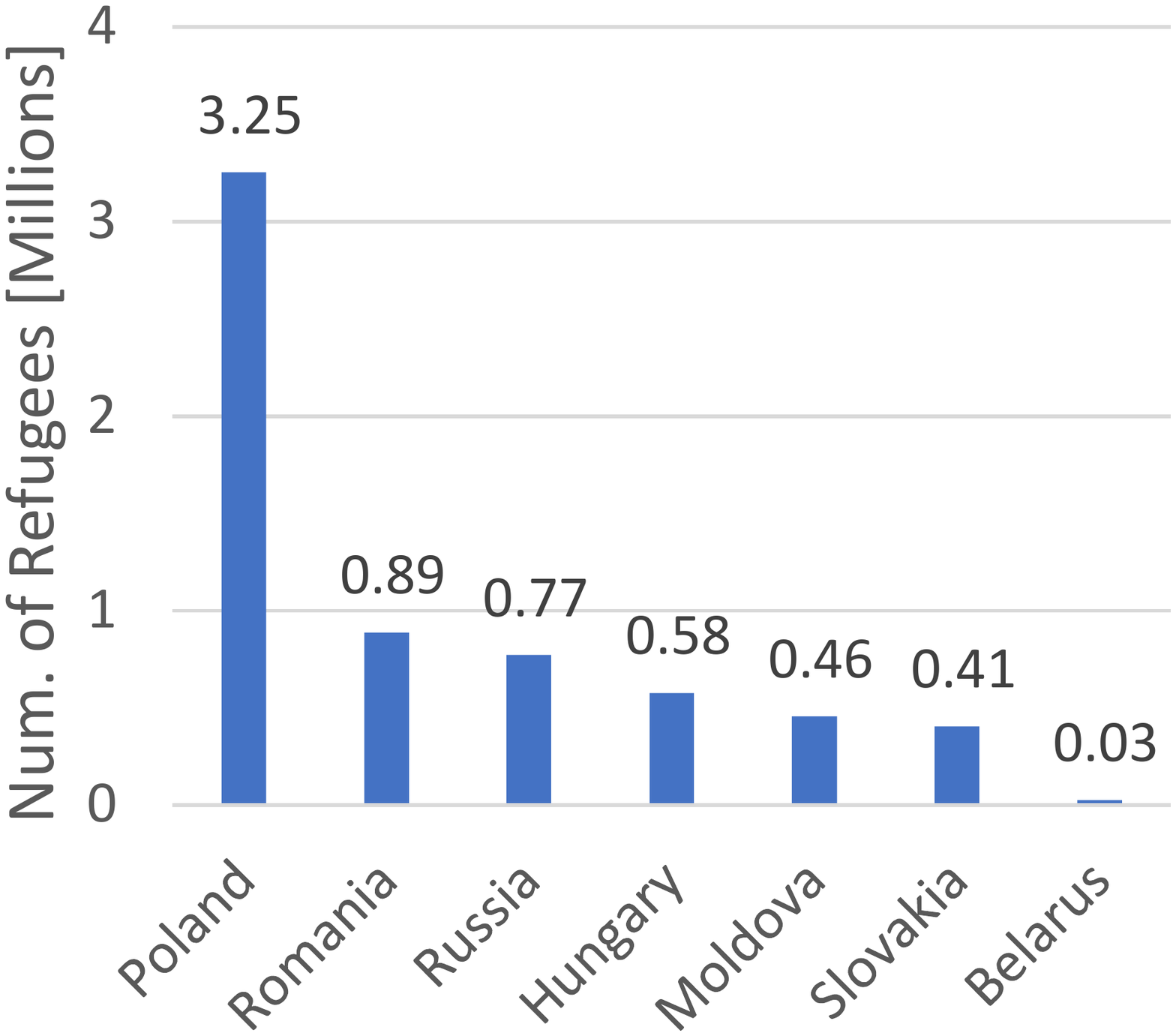}}
  \captionsetup{justification=centering}
  \caption{Number of refugees crossing to each country}
  \label{fig:RefHist}
  \end{subfigure}%
  \caption{Accumulated number of refugees [UNHCR]}
  \label{fig:RefRefHist}
\end{figure}

Preliminary findings about the impact of the war on the Internet performance in Ukraine have been discussed in~\cite{Resilience,UkrainianInternet}. 
\ifdefined\TechReport
The dramatic differences between the Internet performance changes in Ukraine and those in Russia at the beginning of the conflict are presented in~\cite{Asymmetric}.
\fi
Clearly the connectivity during the war was affected by infrastructure aspects such as power outages or damaged communication lines and equipment, causing routing changes and congestion along bottlenecks, and in some cases resulting in user performance degradation. We argue that the refugee crisis was also a major factor in some of these performance changes.

In this paper we use Internet measurement data from multiple sources to show that the Ukrainian refugee crisis affected specific Internet performance metrics not only in Ukraine, but in other countries as well. We demonstrate that during the first 2-3 weeks of the conflict various performance metrics such as the average traffic rate and the mobile device usage significantly changed, having a clear correlation to the flow of refugees.

Our analysis shows how publicly available Internet measurement data can be used to analyze the geographic distribution and the flow of refugees over time. We use website analytics of Ukrainian sites in order to provide a maximum likelihood estimation of the Ukrainian presence in each country.
Our analysis is based solely on publicly available information, and produces large-scale statistics that do not compromise privacy aspects. 
We believe that our approach can be used to aid the ongoing effort in mapping the refugee crisis in order to help and support the refugees. 

\begin{figure*}[htbp]
  \centering
  \begin{subfigure}[t]{.24\textwidth}
  \centering
  \fbox{\includegraphics[trim={6cm 1cm 7cm 1cm},clip,height=6.75\grafflecm]{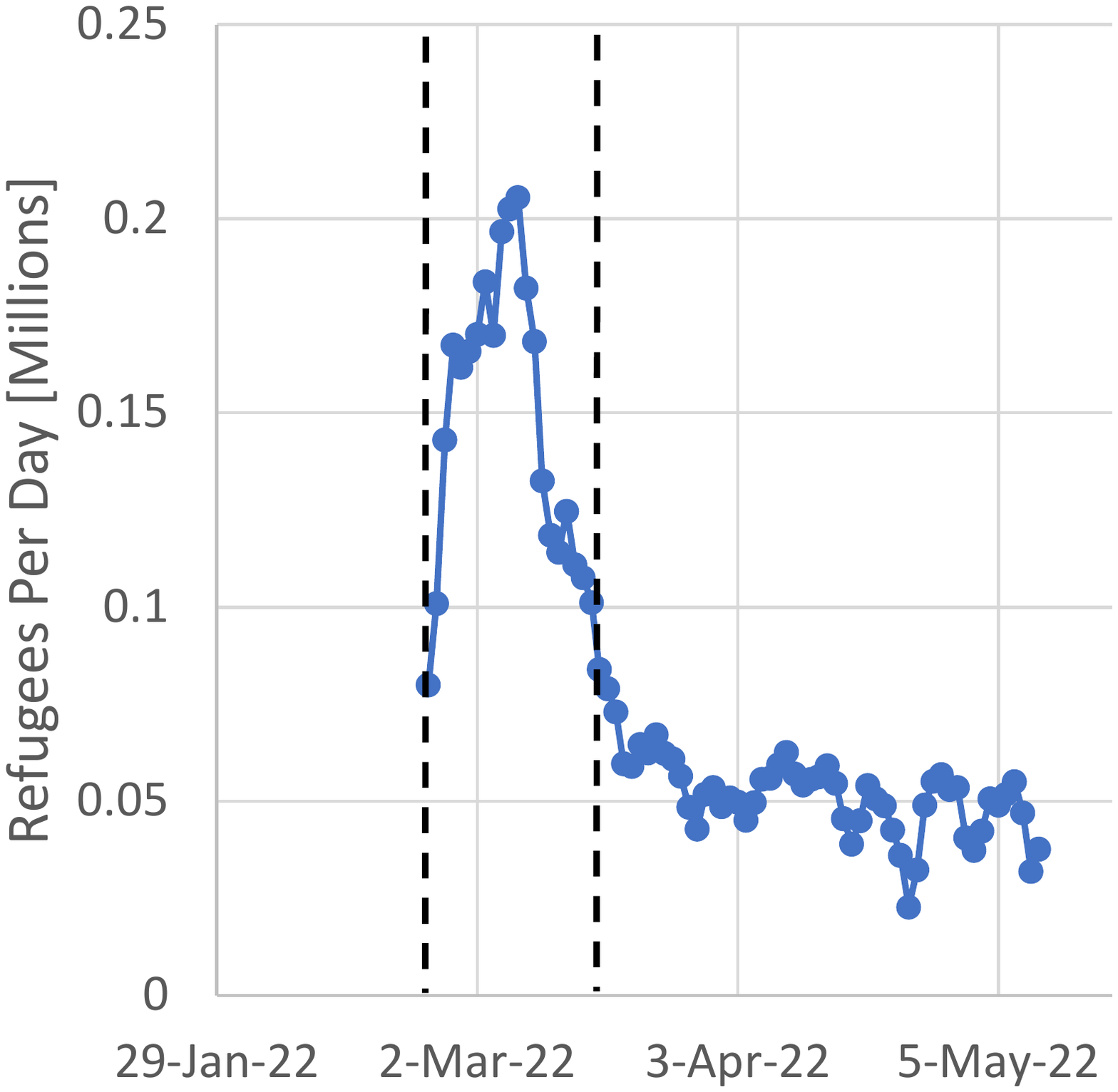}}
	\captionsetup{justification=centering}
  \caption{Refugees exiting Ukraine per day [UNHCR]}
  \label{fig:RefPerDay}
  \end{subfigure}%
  \begin{subfigure}[t]{.24\textwidth}
  \centering
  \fbox{\includegraphics[trim={6cm 1cm 6cm 1cm},clip,height=6.75\grafflecm]{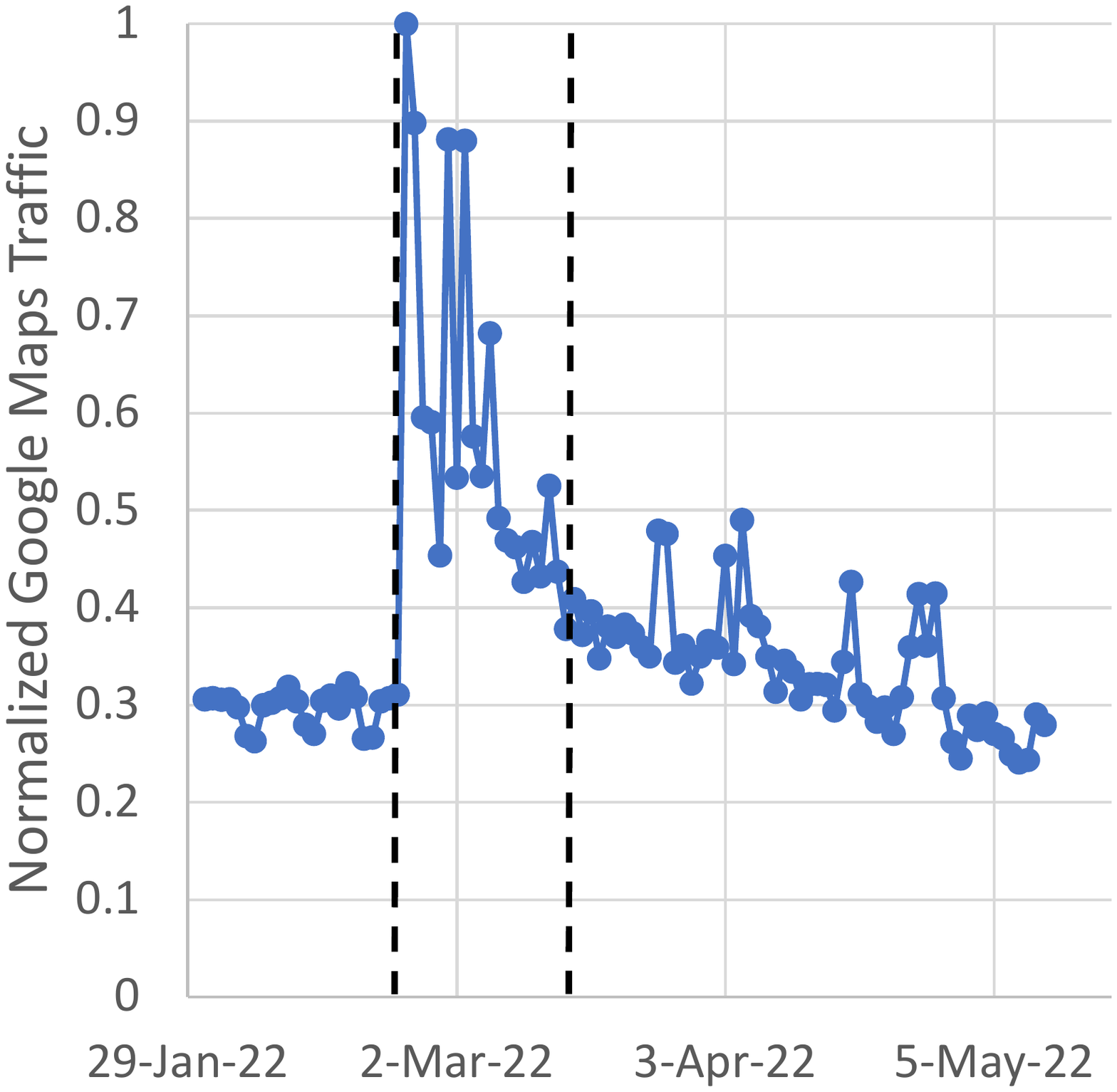}}
  \captionsetup{justification=centering}
  \caption{Normalized Google Maps traffic rate}
  \label{fig:UkrGoogleMaps}
  \end{subfigure}%
  \begin{subfigure}[t]{.24\textwidth}
  \centering
  \fbox{\includegraphics[trim={5cm 1cm 7cm 1cm},clip,height=6.75\grafflecm]{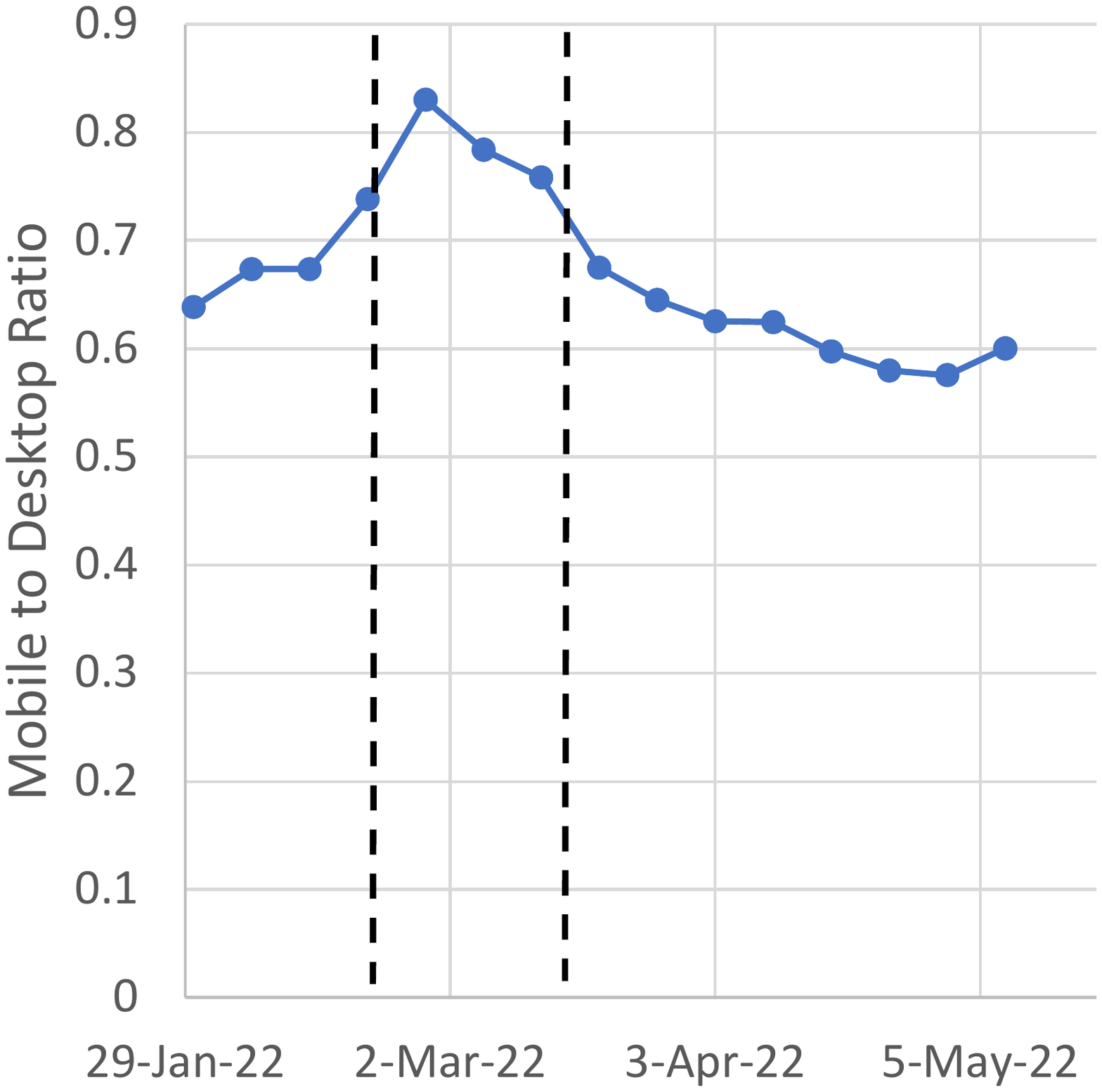}}
	\captionsetup{justification=centering}
  \caption{Mobile to desktop ratio in Ukraine [Statcounter]}
  \label{fig:UkrMobileRatio}
  \end{subfigure}%
  \begin{subfigure}[t]{.24\textwidth}
  \centering
  \fbox{\includegraphics[trim={5cm 1cm 7cm 1cm},clip,height=6.75\grafflecm]{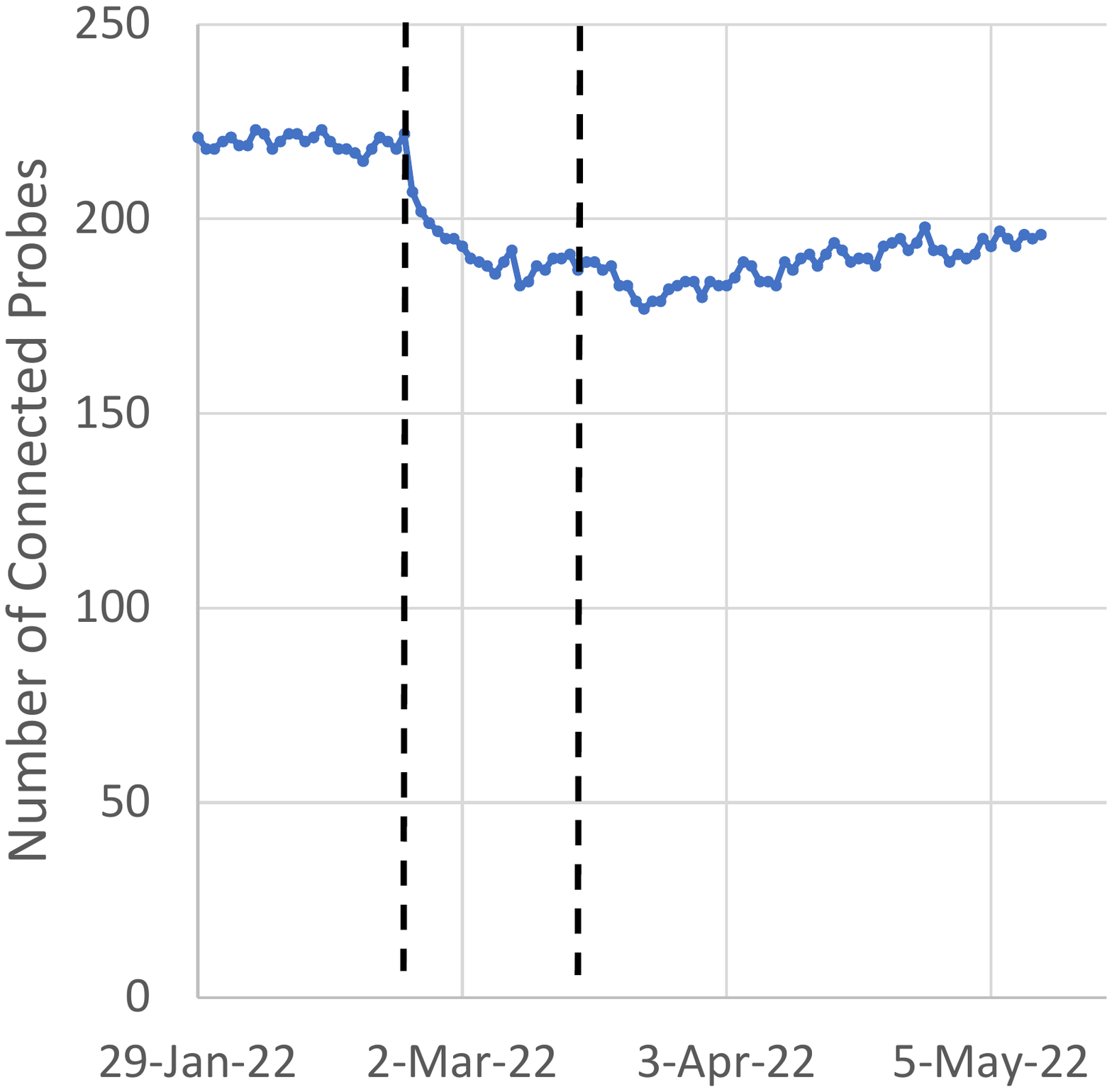}}
	\captionsetup{justification=centering}
  \caption{The number of connected RIPE Atlas probes}
  \label{fig:Connected}
  \end{subfigure}%
  \caption{Refugee flow and its impact. Dotted lines mark the first three weeks.}
  \label{fig:UkrRefMaps}
\end{figure*}

The rest of this paper is organized as follows. Section~\ref{DataSec} describes the sources of data used in this paper. The impact of the refugee crisis on the Internet performance in Ukraine and the neighboring countries is discussed in Section~\ref{UkraineSec}, and Section~\ref{RefSec} introduces a measurement-based mapping of the refugee crisis. 
Related work is discussed in Section\ref{RelWorkSec}, and concluding remarks are presented in Section~\ref{ConcSec}.

\section{Data Sources}
\label{DataSec}

\ifdefined\CutSpace
\textbf{UNHCR.}
\else
\subsection{UNHCR}
\fi
The United Nations High Commissioner for Refugees (UNHCR) publishes daily updates~\cite{UNHCR} about the number of refugees from Ukraine arriving to each of Ukraine's neighboring countries. This data is based on information provided to the UNHCR by authorities from border crossing points. As noted above, while these statistics provide information about Ukrainians crossing the border to neighboring countries, individuals may move freely within European countries, and therefore it is estimated that a large number of people have moved onwards to other countries. The data in this paper was extracted from~\cite{UNHCR} on the second week of May 2022. 

\ifdefined\CutSpace
\textbf{Statcounter.}
\else
\subsection{Statcounter}
\fi
Statcounter~\cite{Statcounter} is a web analytics tool that tracks information about web page views collected from over 2 million sites globally. Statcounter allows flexible extraction of data on a per-country basis including, for example, web browser or social network statistics. The statistics we used in the current work include mobile and desktop usage rates, mobile vendor usage rates, and web search engine rates. Most of this data was analyzed over a period of 3 months, with some of the data analyzed over a period of several years.

\ifdefined\CutSpace
\textbf{Similarweb.}
\else
\subsection{Similarweb}
\fi
Similarweb~\cite{Similarweb} publishes various website analytics data. In the current paper we used Similarweb's ranking of the 50 top Ukrainian websites, as well as the number of visits per month to each website we analyzed. This data was accessed on 16 May 2022.

\ifdefined\CutSpace
\textbf{Google Transparency Report.}
\else
\subsection{Google Transparency Report}
\fi
Google's transparency report~\cite{Google} includes continuously updated data about traffic rates. The site provides data on a per-country basis for each of Google's products such as Web Search and YouTube. The data provides the ratio between the country's request rate and the worldwide request rate for each of Google's products. In this paper we used data over a period of 3 months. For each day in the measurement period we extracted the peak normalized traffic rate, allowing to analyze the trends of the normalized traffic rate in each of the countries we analyzed.

\ifdefined\CutSpace
\textbf{Cloudflare Radar.}
\else
\subsection{Cloudflare Radar}
\fi
Cloudflare Radar~\cite{Cloudflare} is a site that presents detailed data about Cloudflare's traffic. The site enables per-country filtering, including per-country normalized traffic rate over the last 60 days. We used this sliding window of 60 days to keep track of the traffic rates in Ukraine and six of its neighboring countries for a period of 3 months. For each day during this period we extracted the normalized peak traffic rate. Cloudflare also published website visit rate on a per-country basis, which was used in our analysis of Ukrainian websites. This data was accessed on 16 May 2022.

\ifdefined\CutSpace
\textbf{RIPE Atlas.}
\else
\subsection{RIPE Atlas}
\label{RipeSec}
\fi
RIPE Atlas~\cite{RIPEatlas} is a global measurement platform that uses over 11000 probes spread throughout the world. Measurements are continuously performed and published, as well as the status of each of the probes. We analyzed data about the connectivity status of about 200 probes in Ukraine over a period of a year.

\ifdefined\CutSpace
\textbf{Speedtest.}
\else
\subsection{Speedtest}
\label{SpeedSec}
\fi
Speedtest~\cite{Speedtest} by Ookla is one of the most commonly used sites for web-based Internet performance testing. 
Speedtest result statistics are published monthly~\cite{Speedtest} on a per-country basis, including Speedtest's Global Index ranking; each country is ranked according to the median download speed. A separate ranking is published for fixed and for mobile tests. In this paper we used Speedtest data to analyze how the ranking of Ukraine and each of the neighboring countries changed since the beginning of the conflict.

\section{Impact of the Refugee Crisis on Internet Performance}
\label{UkraineSec}
\subsection{Performance in Ukraine}
As previously discussed, the Internet performance in Ukraine after the beginning of the conflict was affected by infrastructure damage and outages. However, as we show in this section, some of the performance metrics and behavior were significantly different in the first 2-3 weeks of the war due to the large flow of refugees, as well as vast population displacement within Ukraine. 

The rate of refugees exiting Ukraine according to~\cite{UNHCR} is illustrated in Figure~\ref{fig:RefPerDay}, with the first 3 weeks marked by dotted lines, emphasizing the high rate during this period. Figures~\ref{fig:UkrGoogleMaps}-\ref{fig:Connected} illustrate the unusual performance during this period of time, showing a high correlation with the 
\ifdefined\CutSpace
refugee rate. 
\else
refugee rate, which suggests that the most significant factor that affected these performance changes during this 3 week period is the refugee crisis.
\fi
The figures throughout this paper are marked with a dotted line, indicating the beginning of the conflict, and in some cases there are two dotted lines, marking the first 3 weeks of the conflict. 

An unusual trend is shown in Google Maps traffic over the marked 3 week period, depicted in Figure~\ref{fig:UkrGoogleMaps}. The Google Maps traffic rate increased by over 200\% in the first days of the war, due to the wave of refugees and domestic transport, and after three weeks was only 25\% higher than normal, gradually returning to normal over the following two months.

Another aspect that indicates the flow of refugees during these 2-3 weeks is the nature of mobile network usage. Based on Statcounter data we computed the ratio between mobile device traffic and desktop traffic. Figure~\ref{fig:UkrMobileRatio} illustrates the mobile-to-desktop ratio in Ukraine, indicating a steep increase of over 10\% in mobile device usage in the 2-3 weeks after the war started, and a large decrease in the ratio afterwards, which can be explained by the traveling and displacement during the first 2-3 weeks, which resulted in high mobile device usage compared to desktops, followed by a general decrease in mobile usage in the period that followed.

Figure~\ref{fig:Connected} depicts the number of RIPE Atlas~\cite{RIPEatlas} active (connected) probes in Ukraine. 
Specifically, 222 probes were active in Ukraine on 23 February 2022, with an average of 219.9 probes during the year beforehand. The figure shows a steep decline to just 183 probes during the first two weeks of the war, with no significant changes afterwards. This behavior is not likely to be due to connectivity issues, which would likely result in more fluctuations. Instead, this behavior once again indicates a high correlation to the internal displacement and the refugee flow (Figure~\ref{fig:RefPerDay}).

Another aspect that was analyzed is the traffic rate in Ukraine at the beginning of the war and slightly beforehand. As shown in Figure~\ref{fig:UkrTrafficRate}, the traffic rate dropped during the first 2-3 weeks of the war, and then gradually started to increase.

\begin{figure}[htbp]
  \centering
  \begin{subfigure}[t]{.24\textwidth}
  \centering
  \fbox{\includegraphics[trim={6cm 1.2cm 6cm 1cm},clip,height=6.75\grafflecm]{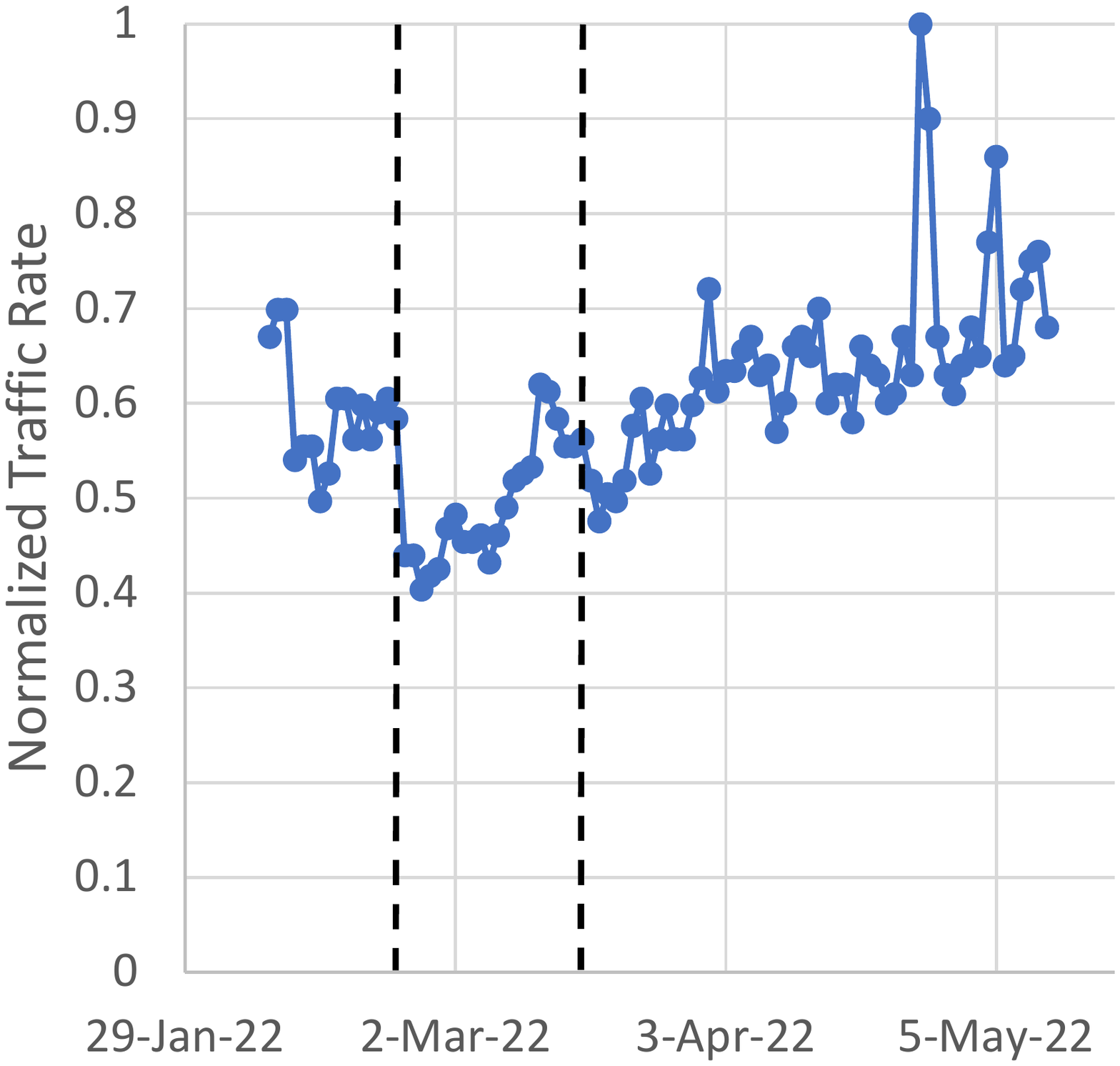}}
  \captionsetup{justification=centering}
  \caption{Normalized traffic [Cloudflare]}
  \label{fig:UkrCloudflare}
  \end{subfigure}%
  \begin{subfigure}[t]{.24\textwidth}
  \centering
  \fbox{\includegraphics[trim={6cm 1cm 6cm 1cm},clip,height=6.75\grafflecm]{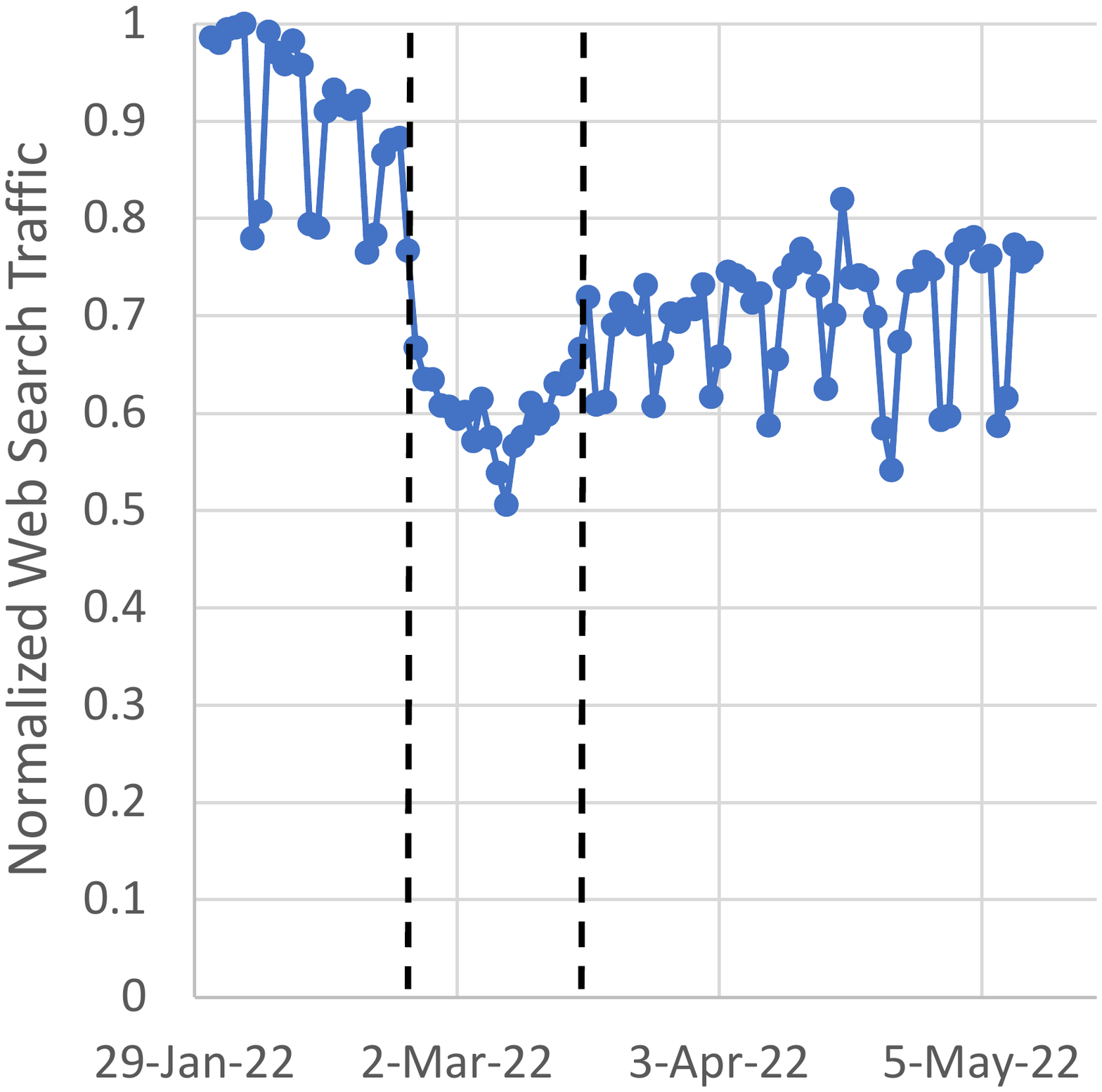}}
  \captionsetup{justification=centering}
  \caption{Normalized web search traffic [Google]}
  \label{fig:UkrGoogleWeb}
  \end{subfigure}%
  \caption{Traffic rate in Ukraine}
  \label{fig:UkrTrafficRate}
\end{figure}

We first observe the normalized daily peak traffic rate in Ukraine, measured by Cloudflare, as shown in Figure~\ref{fig:UkrCloudflare}. We found that the peak traffic rate in the two weeks after the war began was about 25\% lower than during the two weeks beforehand. A similar trend is shown in Google's data, depicted in Figure~\ref{fig:UkrGoogleWeb}; the normalized daily peak web search traffic rate decreased by 30\% during the same period of time.

These low traffic rates lasted for 2-3 weeks and then gradually climbed back to the normal rates, a trend that can be explained by the refugee crisis and the internal population displacement, having a temporary effect that lasted until the displaced population managed to settle in.  

An analysis of the types of mobile devices used in Ukraine during the first few weeks of the war reveals a steep change in the mobile device vendors used in Ukraine. The mobile vendor distribution in previous years, depicted in Figure~\ref{fig:UkrMobileLong}, focuses on four of the most common mobile device vendors during this period, and shows the gradual decrease in Nokia mobile devices, from 13\% in 2015 to under 1\% in 2021. However, during the first weeks of the war in 2022, as shown in Figure~\ref{fig:UkrMobileShort}, the rate of Nokia device usage increased back to 13\%. This surprising change is explained by the increasing demand for mobile devices, causing people to revive older and unused mobile devices. This trend gradually decreased during the first two months of the war.

\ifdefined\UpDownFigures
\begin{figure}[!b]
\else
\begin{figure}[htbp]
\fi
  \centering
  \begin{subfigure}[t]{.24\textwidth}
  \centering
  \fbox{\includegraphics[trim={6cm 3cm 7cm 2cm},clip,height=6\grafflecm]{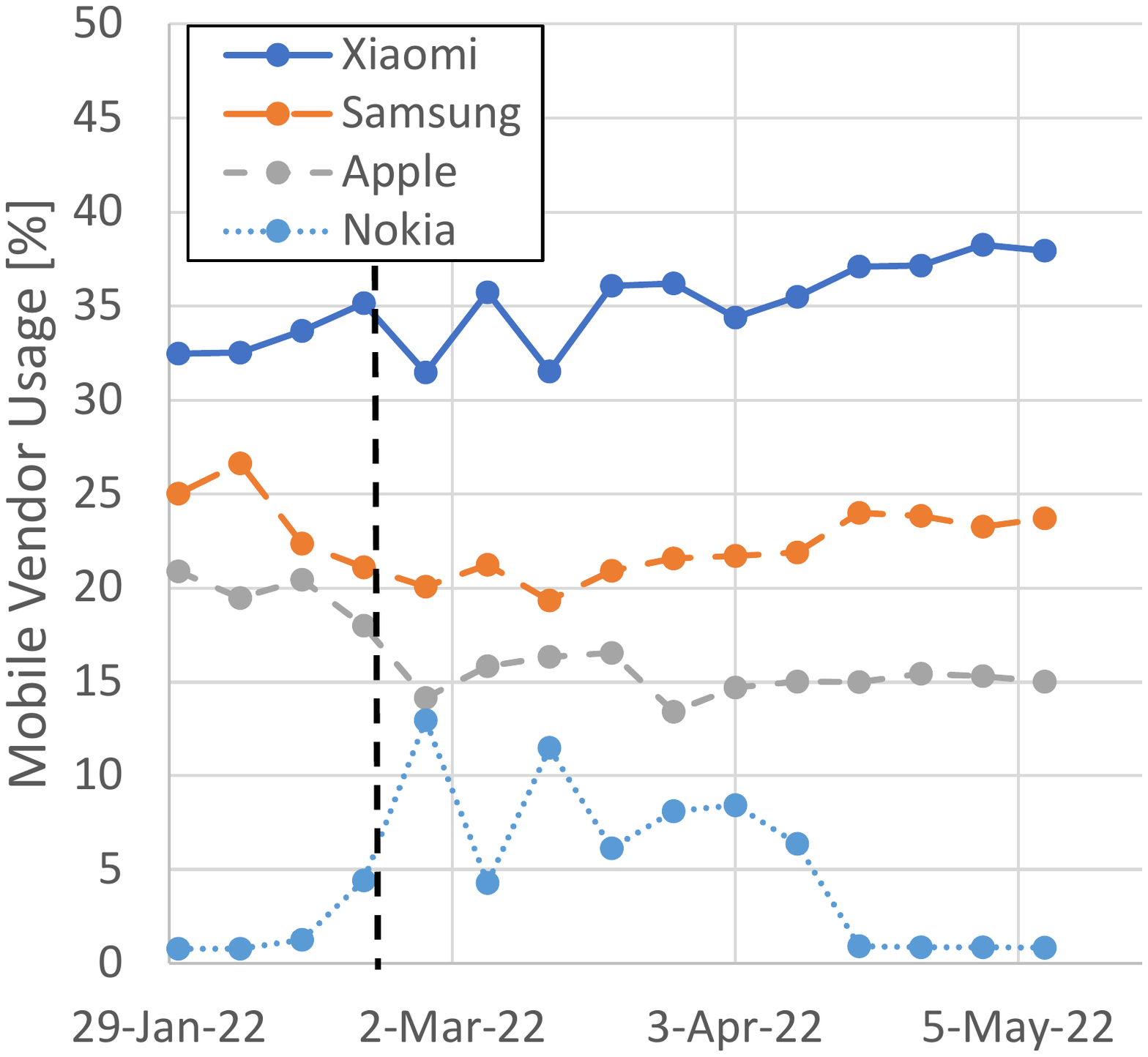}}
  \captionsetup{justification=centering}
  \caption{At the beginning of 2022}
  \label{fig:UkrMobileShort}
  \end{subfigure}%
  \begin{subfigure}[t]{.24\textwidth}
  \centering
  \fbox{\includegraphics[trim={6cm 3cm 7cm 2cm},clip,height=6\grafflecm]{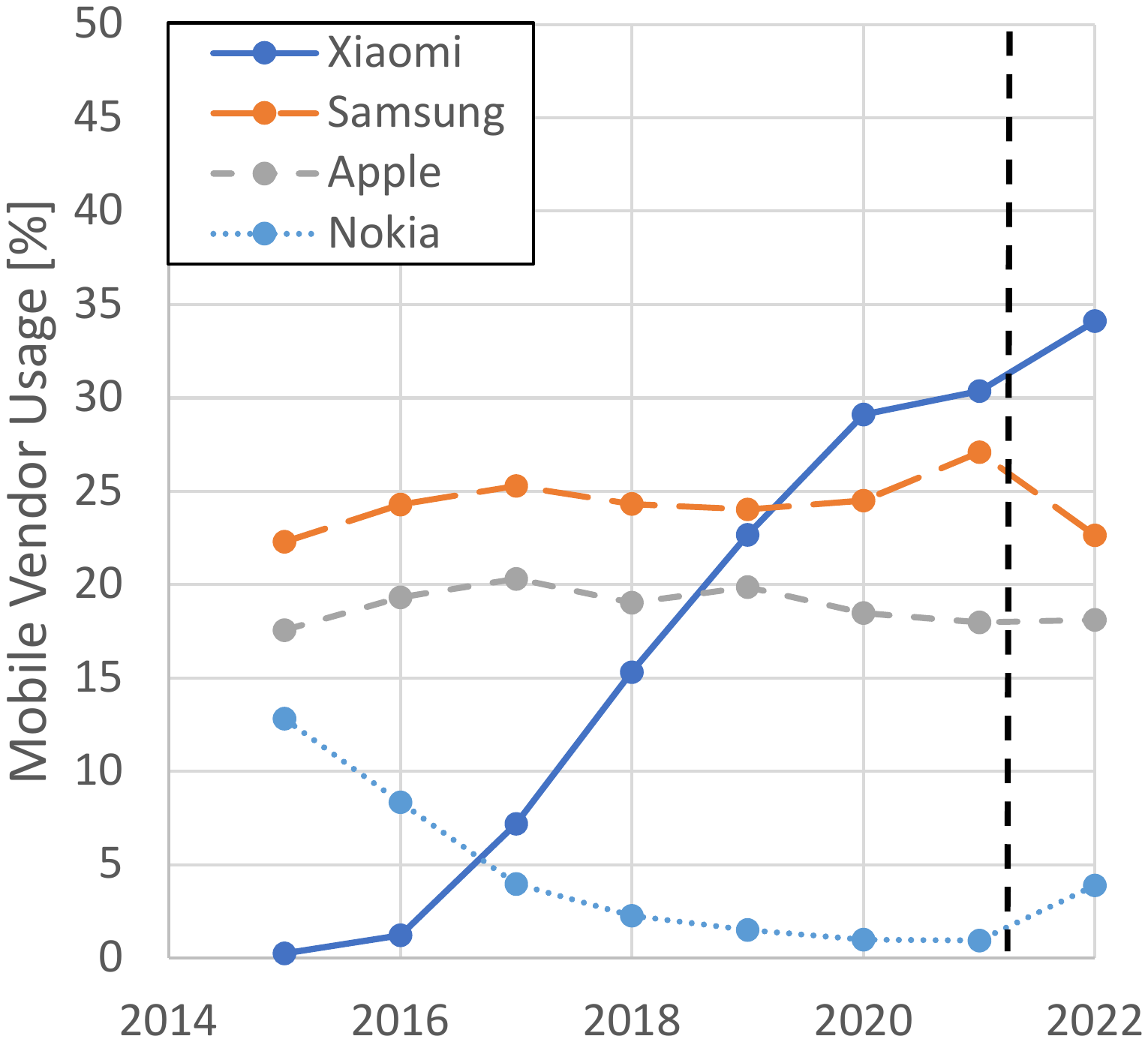}}
  \captionsetup{justification=centering}
  \caption{Previous years}
  \label{fig:UkrMobileLong}
  \end{subfigure}%
  \caption{Mobile vendor usage in Ukraine [Statcounter]}
  \label{fig:UkrMobile}
\end{figure}

\subsection{Performance around Ukraine}
One of the aspects we analyzed is the impact of the crisis on the Internet performance of countries around Ukraine.  
Since Poland was the country that received the largest number of refugees (Figure~\ref{fig:RefHist}), we take Poland as a case study, and focus our analysis in the current subsection on Poland.

We start by analyzing Internet access performance, based on Speedtest data, which shows the rank of each country for fixed networks and for mobile networks. Based on these measurements, we found that following the beginning of the war, four of the seven countries surrounding Ukraine were ranked lower than before the beginning of the war. 
Specifically, Poland was ranked lower in March 2022 than in February, both in fixed (one place lower) and in mobile tests (two places lower).

\ifdefined\UpDownFigures
\begin{figure}[!t]
\else
\begin{figure}[htbp]
\fi
  \centering
  \begin{subfigure}[t]{.25\textwidth}
  \centering
  \fbox{\includegraphics[trim={3.5cm 2cm 7.5cm 1cm},clip,height=6\grafflecm]{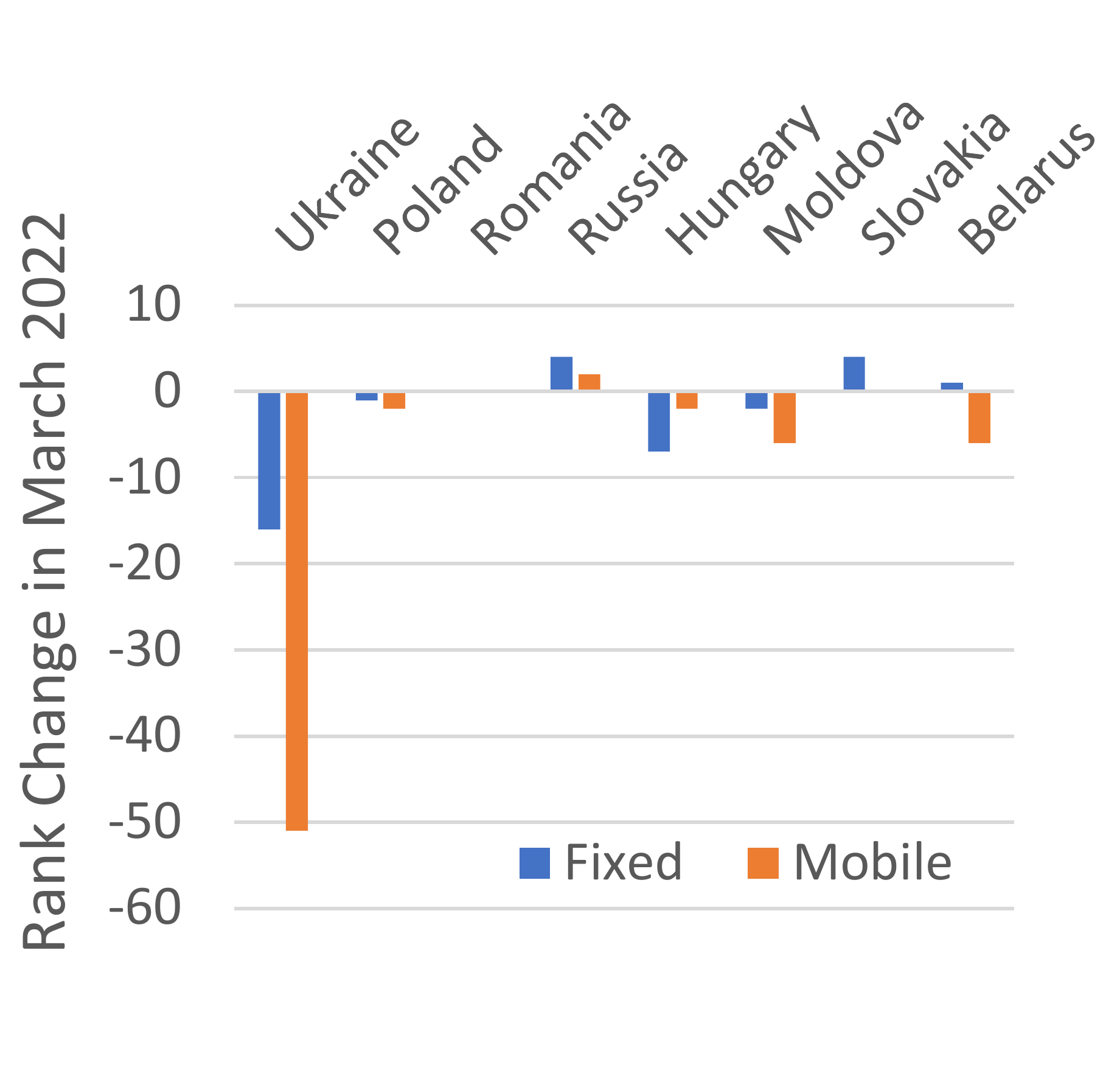}}
  \captionsetup{justification=centering}
  \caption{Speedtest performance rank change in March 2022}
  \label{fig:Rank}
  \end{subfigure}%
  \begin{subfigure}[t]{.25\textwidth}
  \centering
  \fbox{\includegraphics[trim={4cm 0 3cm 0},clip,height=6\grafflecm]{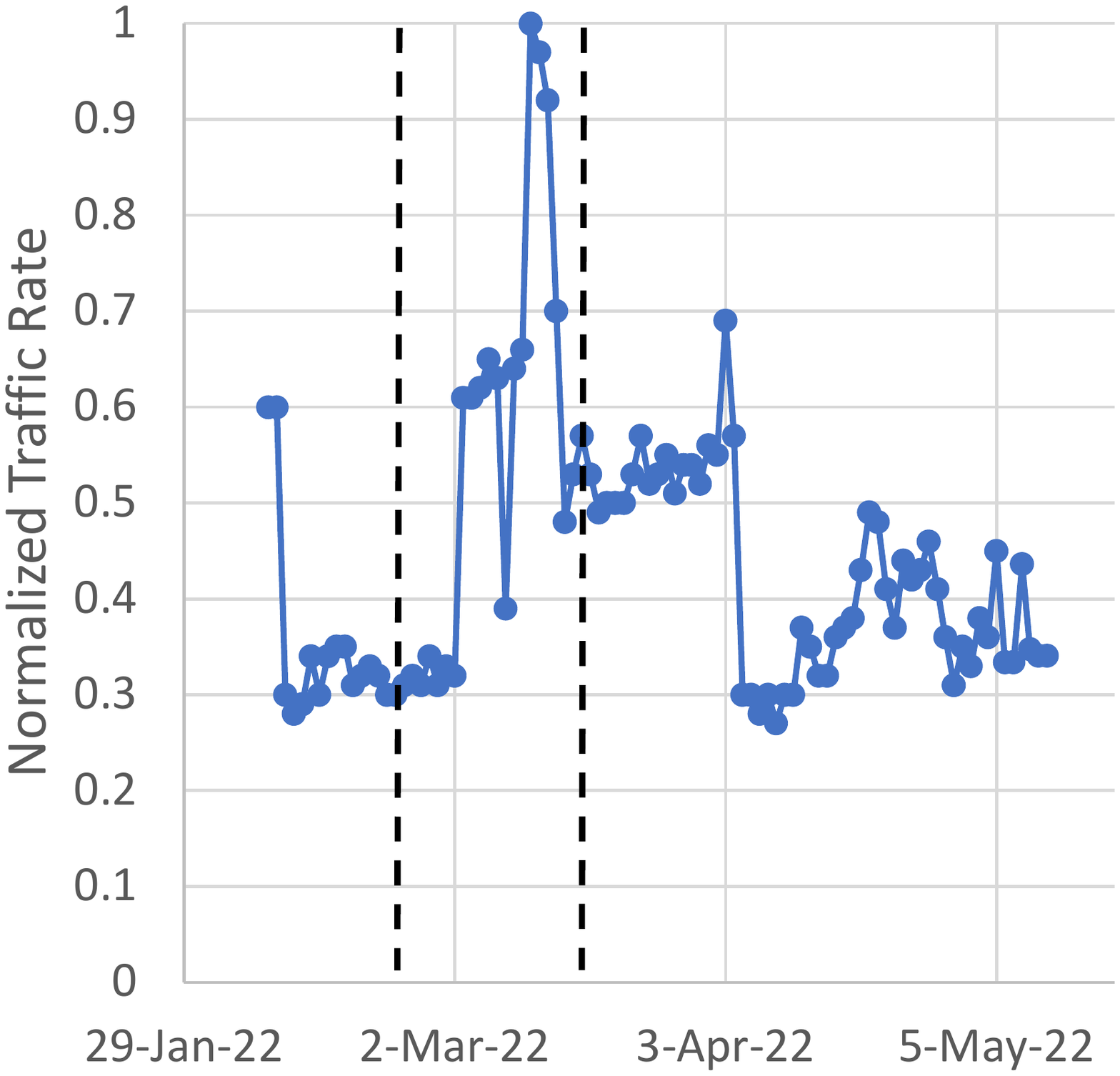}}
  \captionsetup{justification=centering}
  \caption{Poland: normalized traffic rate [Cloudflare]}
  \label{fig:PolandCloudflare}
  \end{subfigure}%
  \caption{Performance impact on neighboring countries}
  \label{fig:NeighborImpact}
\end{figure}

The traffic rate in Poland is another interesting metric in our analysis. We found that the Cloudflare traffic rate, shown in Figure~\ref{fig:PolandCloudflare}, increased by over 40\% in the two weeks after the war started compared to the two weeks beforehand.

\ifdefined\UpDownFigures
\begin{figure}[!b]
\else
\begin{figure}[htbp]
\fi
  \centering
  \begin{subfigure}[t]{.25\textwidth}
  \centering
  \fbox{\includegraphics[trim={5cm 1cm 6cm 2cm},clip,height=6\grafflecm]{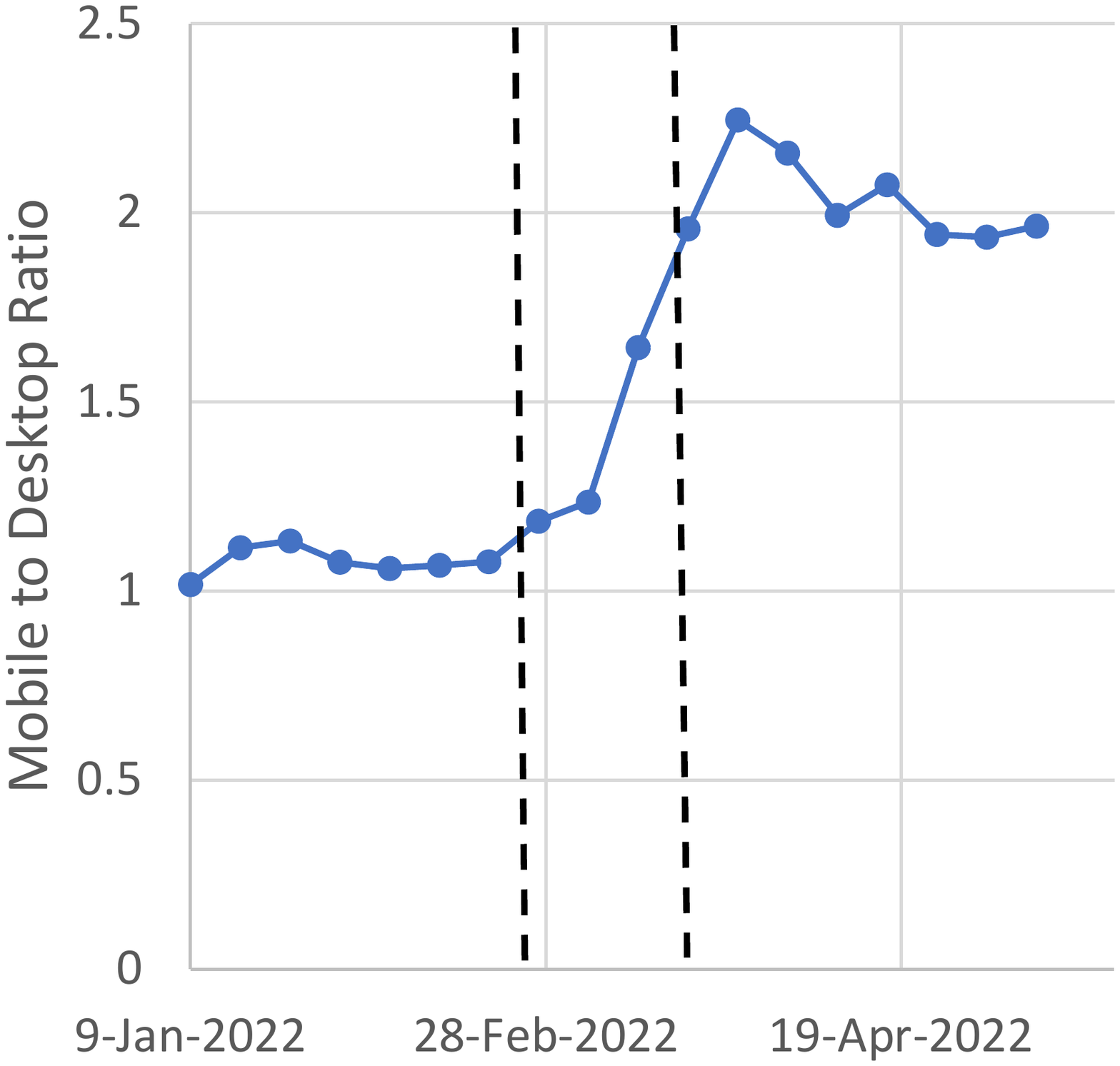}}
	\captionsetup{justification=centering}
  \caption{Mobile to desktop ratio}
  \label{fig:PolMobileRatio}
  \end{subfigure}%
  \begin{subfigure}[t]{.25\textwidth}
  \centering
  \fbox{\includegraphics[trim={5cm 3cm 7cm 2cm},clip,height=6\grafflecm]{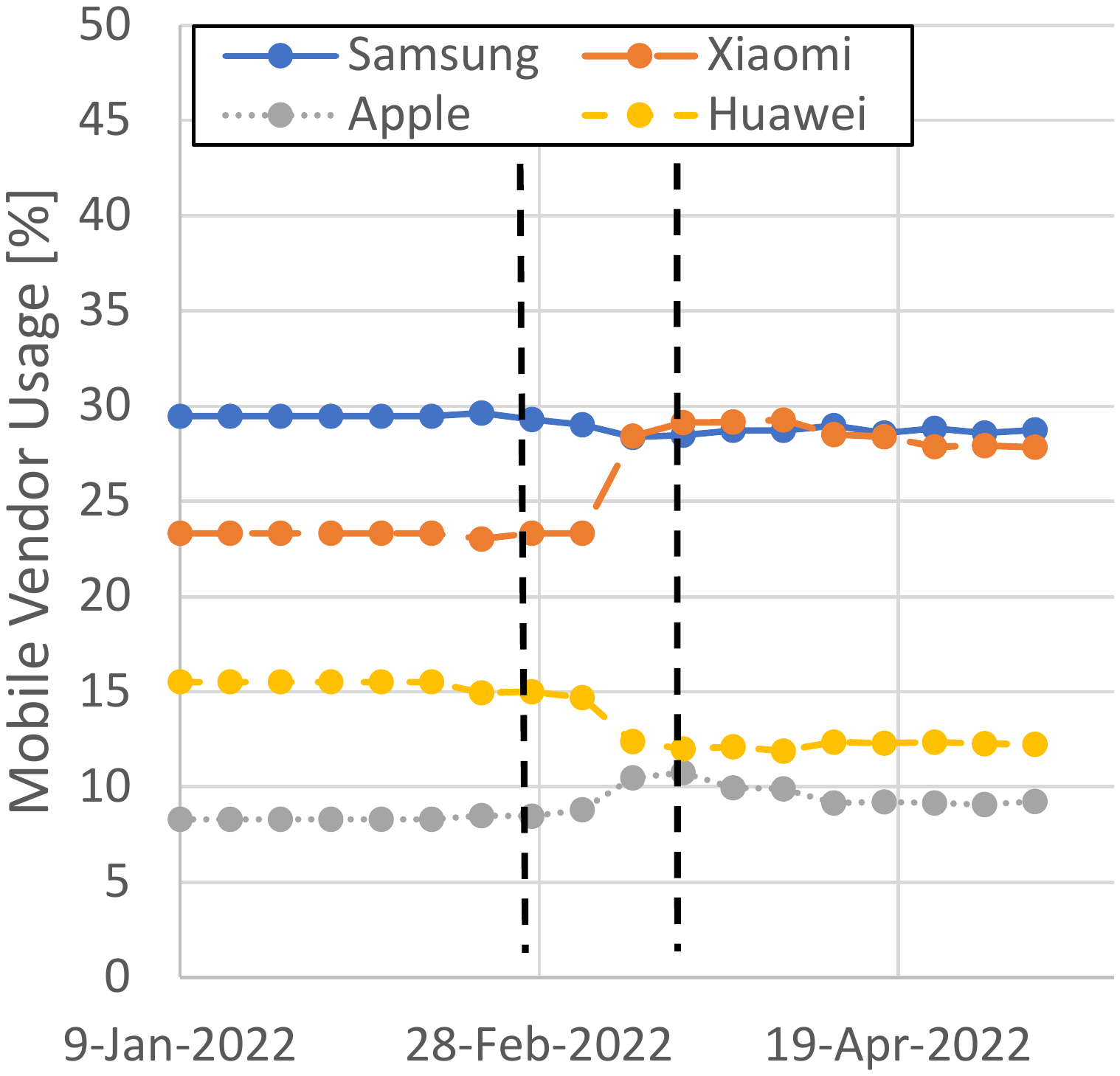}}
  \captionsetup{justification=centering}
  \caption{Mobile vendor usage}
  \label{fig:PolVendor}
  \end{subfigure}%
  \caption{Impact on mobile usage in Poland [Statcounter]}
  \label{fig:PolMobile}
\end{figure}

The mobile-to-desktop ratio in Poland, depicted in Figure~\ref{fig:PolMobileRatio}, demonstrates a significant increase in mobile usage in Poland shortly after the war started, which can be explained by the large number of people who were in transit and were mostly using mobile devices. This assessment is further confirmed by analyzing the mobile vendor usage, as shown in Figure~\ref{fig:PolVendor}.
As previously shown above (Figure~\ref{fig:UkrMobile}), the most commonly used mobile vendor in Ukraine is Xiaomi. As illustrated in Figure~\ref{fig:PolVendor}, the usage rate of Xiaomi devices has significantly increased in Poland during the first 2-3 weeks of the war, and then remained stable. 

The mobile usage in Poland, as shown in Figure~\ref{fig:PolMobile} suggests that the Ukrainian presence in Poland significantly increased during the first 3 weeks of the war, but then remained roughly stable afterwards despite the continued influx (Figure~\ref{fig:Ref}), which confirms the fact that many of the refugees entering Poland continued onward to other countries.

\ifdefined\TechReport
\begin{figure}[htbp]
  \centering
  \fbox{\includegraphics[trim={6cm 1cm 5cm 1cm},clip,height=6\grafflecm]{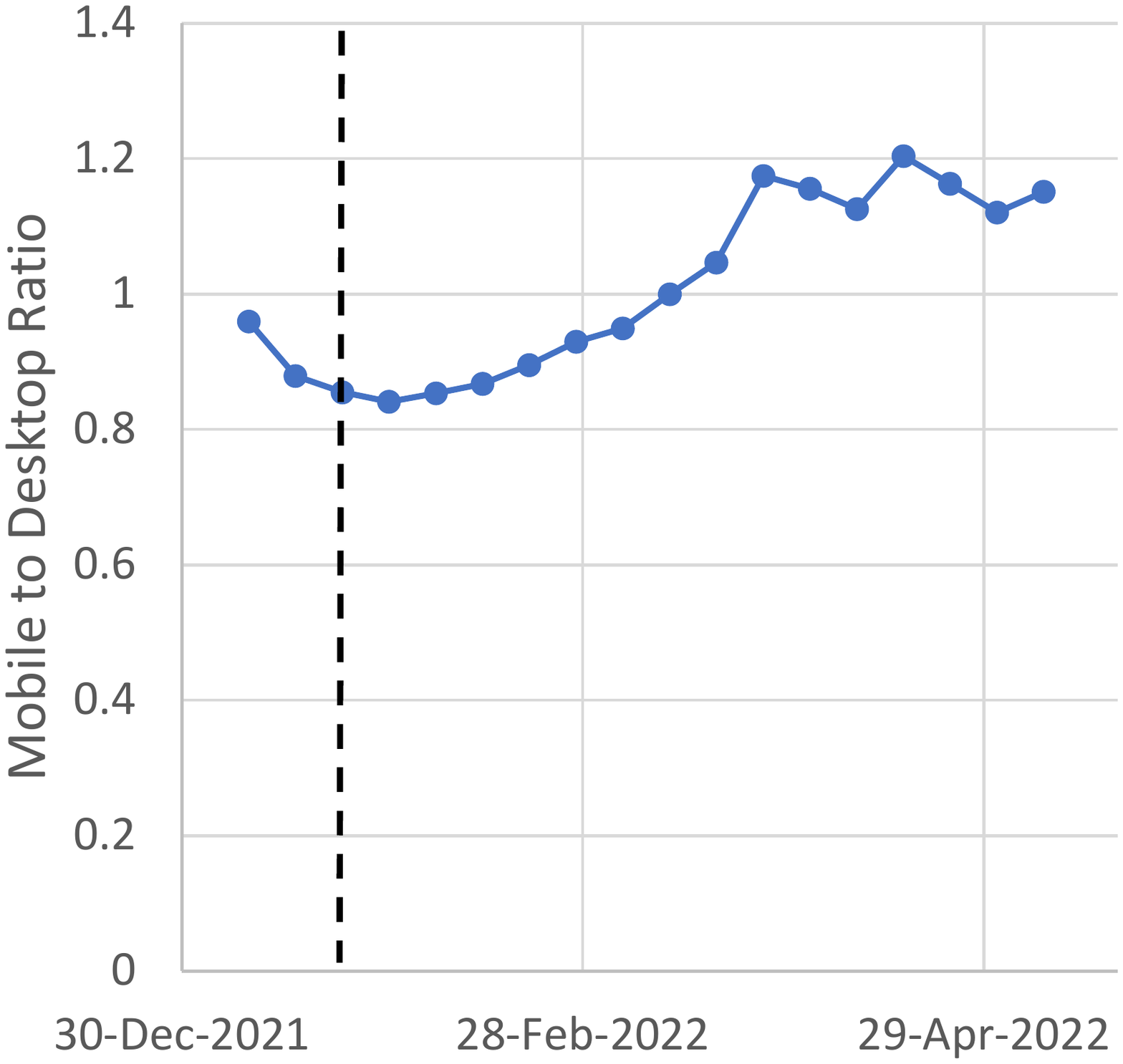}}
  \captionsetup{justification=centering}
  \caption{Mobile to desktop ratio in Europe. The dotted line marks the beginning of the removal of COVID-19 travel restrictions.}
  \label{fig:EuropeMobile}
\end{figure}
\fi

\begin{figure*}[htbp]
	\centering
  \fbox{\includegraphics[trim={2cm 8.5cm 2cm 7.5cm},clip,height=5\grafflecm]{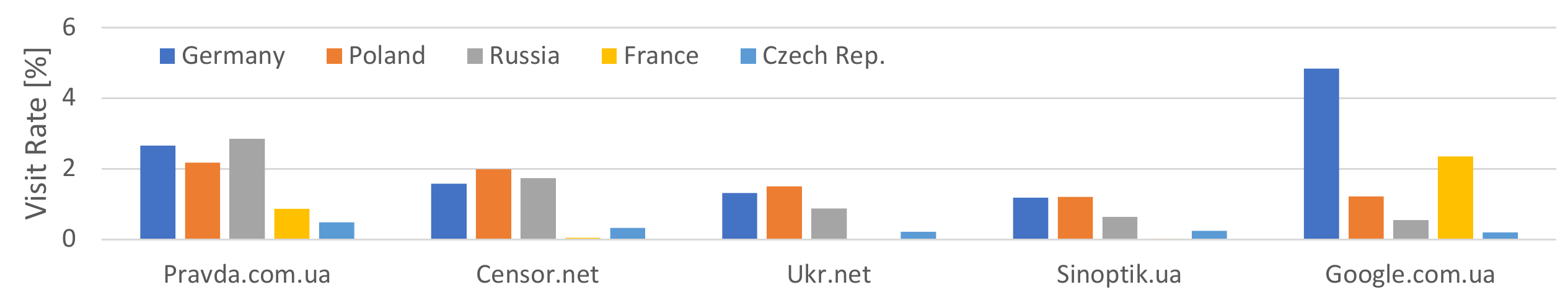}}
	\captionsetup{justification=centering}
  \caption{Visit rate to the top five Ukrainian websites from five countries in Europe [Cloudflare]}
  \label{fig:SiteHist}
\end{figure*}

\ifdefined\TechReport
These results show that measurement data about mobile device usage provides a clear indication about refugee presence in Poland, which received the largest number of refugees. However, this approach was not necessarily productive when we analyzed other countries in Europe. One of the important trends we noticed is the fact that starting from mid-January the mobile-to-desktop ratio has increased throughout Europe, following the removal of the COVID-19 travel restrictions~\cite{DW} that were applied during the Omicron wave. With the travel restrictions removed, mobile traffic increased significantly with increasingly growing numbers of people traveling and using mobile devices. The high mobile network usage (Figure~\ref{fig:PolMobile}) in Poland was clearly due to the refugee wave, indicated by the date in which the steep increase started. However, when we tried to analyze mobile device usage in the rest of Europe, we found that it was deeply affected by the removal of the COVID restrictions, as shown in Figure~\ref{fig:EuropeMobile}, also causing significant change in mobile vendor usage.
\fi

The Poland case study shows that the refugee crisis affected not only the performance in Ukraine, but also in other countries. However, it should be noted that the indications we presented in the current subsection were less dominant in other countries we analyzed, and therefore these metrics would not be a practical means for mapping the refugees' presence. However, the request rate to Ukrainian websites was found to be an effective metric for this purpose, as further discussed below.

\section{Mapping The Refugee Crisis}
\label{RefSec}

Many of the refugees who crossed the border to neighboring countries continued their journey to other countries. While the UNHCR keeps track of the number of refugees that have reached each of the neighboring countries, there is no accurate record of the number of refugees who stayed in these countries, or generally the number of refugees currently staying in each country in the world. One of the keys to understanding the crisis and helping the refugees is being able to assess how many refugees are staying in each country.

\subsection{A Geographical Perspective}
One of the main approaches we analyzed in order to map the Ukrainian presence in the world is to observe the visit rate to top Ukrainian websites from locations around the world. We crossed information from two sources: the top Ukrainian sites and the number of visits per month were extracted from Similarweb~\cite{Similarweb}, and the per-country visit rates for these websites are based on data from Cloudflare~\cite{Cloudflare}. 

The visit rate to the top five Ukrainian websites is illustrated in Figure~\ref{fig:SiteHist}, focusing on the five countries in Europe which had the highest visit rates (excluding Ukraine). For example, 4.84\% of the visits to google.com.ua came from Germany.

In our analysis we used data from 15 sites in order to estimate the Ukrainian presence in each country. These sites were taken from the list of top Ukrainian sites~\cite{Similarweb}, eliminating international sites such as facebook.com and yandex.ru, and focusing on sites that are predominantly accessed by Ukrainian users.

We now present a brief overview of the model and assumptions that were used in our estimation method. We denote the total number of countries by $N$, and the total number of Ukrainian website users, spread throughout the world, is denoted by $S$. The number of individuals in each country is denoted by $n_1, n_2, \ldots, n_N$, where the index indicates the country. We assume that every website visit has an equal probability of being accessed by each individual, and therefore the probability of being accessed from country $i$ is $p_i = n_i / S$, and thus we are analyzing a multinomial distribution. We would like to estimate the values $p_i$ for $i=1,\ldots,N$, representing the proportional part of the population in each country $i$. 

For a given set of measurements which includes $A$ website visits, with $x_1, \ldots, x_N$ specifying the number of visits from each country, we can evaluate the estimated values ${\hat p}_i$ by using a Maximum Likelihood (ML) estimator for multinomial distribution (e.g.,~\cite{MLE}), which is given by ${\hat p}_i = x_i / A$.

We analyzed $K$ websites ($K=15$ in our analysis), where for each site $j$ the number of visits per month, $A_j$, was extracted from~\cite{Similarweb}. For each site $j$ we used the values $x_{i,j}$ from~\cite{Cloudflare}, specifying the relative number of visits from country $i$ to site $j$, and thus $x_{i,j} \cdot A_j$ is the absolute number of visits per month. Hence, for each country $i$ we have $x_i = \sum_{j=1}^K x_{i,j} \cdot A_j$. Thus, the ML estimator is as follows:

\begin{figure*}[htbp]
	\centering
  \fbox{\includegraphics[trim={2cm 6.75cm 2cm 7.25cm},clip,height=6\grafflecm]{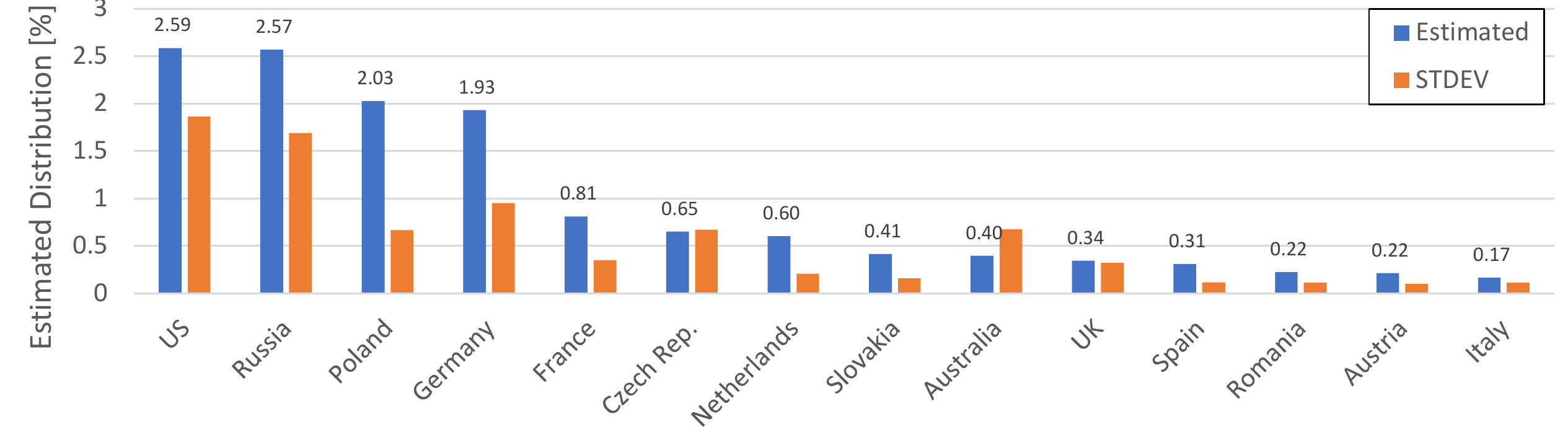}}
	\captionsetup{justification=centering}
  \caption{Estimated Ukrainian presence (percentage from the total Ukrainian population in the world) in foreign countries based on the ML estimation}
  \label{fig:EstimatedHist}
\end{figure*}

\begin{equation}
\label{eq:est}
{\hat p}_i = \frac{\sum_{j=1}^K x_{i,j} \cdot A_j}{\sum_{j=1}^K A_j}
\end{equation}

The results of the estimation are illustrated in Figure~\ref{fig:EstimatedHist}, showing for each of the top countries the estimated number of Ukrainians, expressed as the percentage from the world's Ukrainian population.\footnote{Prior to the war the population in Ukraine was about 44 million~\cite{Worldometer}, and the Ukrainian diaspora was about 6 million~\cite{UNHCR}, summing up to a total of about 50 million.}  For example, it is estimated that 2.03\% of the Ukrainian population is located in Poland, which is about 1 million people. The histogram presents the estimated percentage, computed based on the ML estimator of Eq.~\ref{eq:est}, as well as the standard deviation (STDEV) compared to the estimated value for each country. We note that the STDEV is high for Russia, indicating a low confidence level. However, the estimated presence in Russia is high, indicating over 1 million people, suggesting that the number reported by UNHCR, 0.77 million (Figure~\ref{fig:RefHist}), possibly does not reflect the full picture, and potentially confirms the Russian announcement about transferring over 1 million Ukrainians~\cite{RusRef} to Russia.
Low STDEV values were computed for Germany and Poland and other European countries, showing a higher confidence level in our results.

This analysis can be used as complementary data to the daily UNHCR data of Ukrainian border crossing. Notably, our analysis indicates that there is significant presence of Ukrainians in the US, Germany, France, the Czech Rep. and the Netherlands, as well as other European countries, and not only in Ukraine's neighboring countries.

\ifdefined\TechReport
Focusing on google.com.ua as an example of one of the top websites, Figure~\ref{fig:GoogleHist} illustrates a detailed comparison of the visit rate in various countries in Europe.

\begin{figure*}[htbp]
	\centering
  \fbox{\includegraphics[trim={2cm 7.5cm 2cm 7.5cm},clip,height=6\grafflecm]{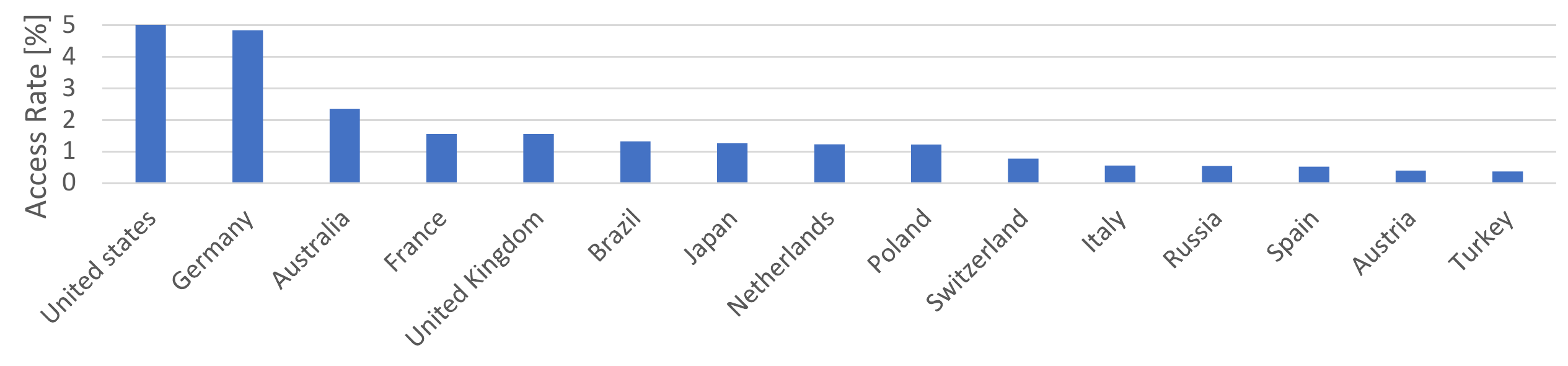}}
	\captionsetup{justification=centering}
  \caption{Visit rate to google.com.ua from various countries in Europe}
  \label{fig:GoogleHist}
\end{figure*}
\fi

\begin{figure*}[htbp]
  \centering
  \begin{subfigure}[t]{.21\textwidth}
  \centering
  \fbox{\includegraphics[trim={3cm 2cm 3.5cm 2cm},clip,height=4\grafflecm]{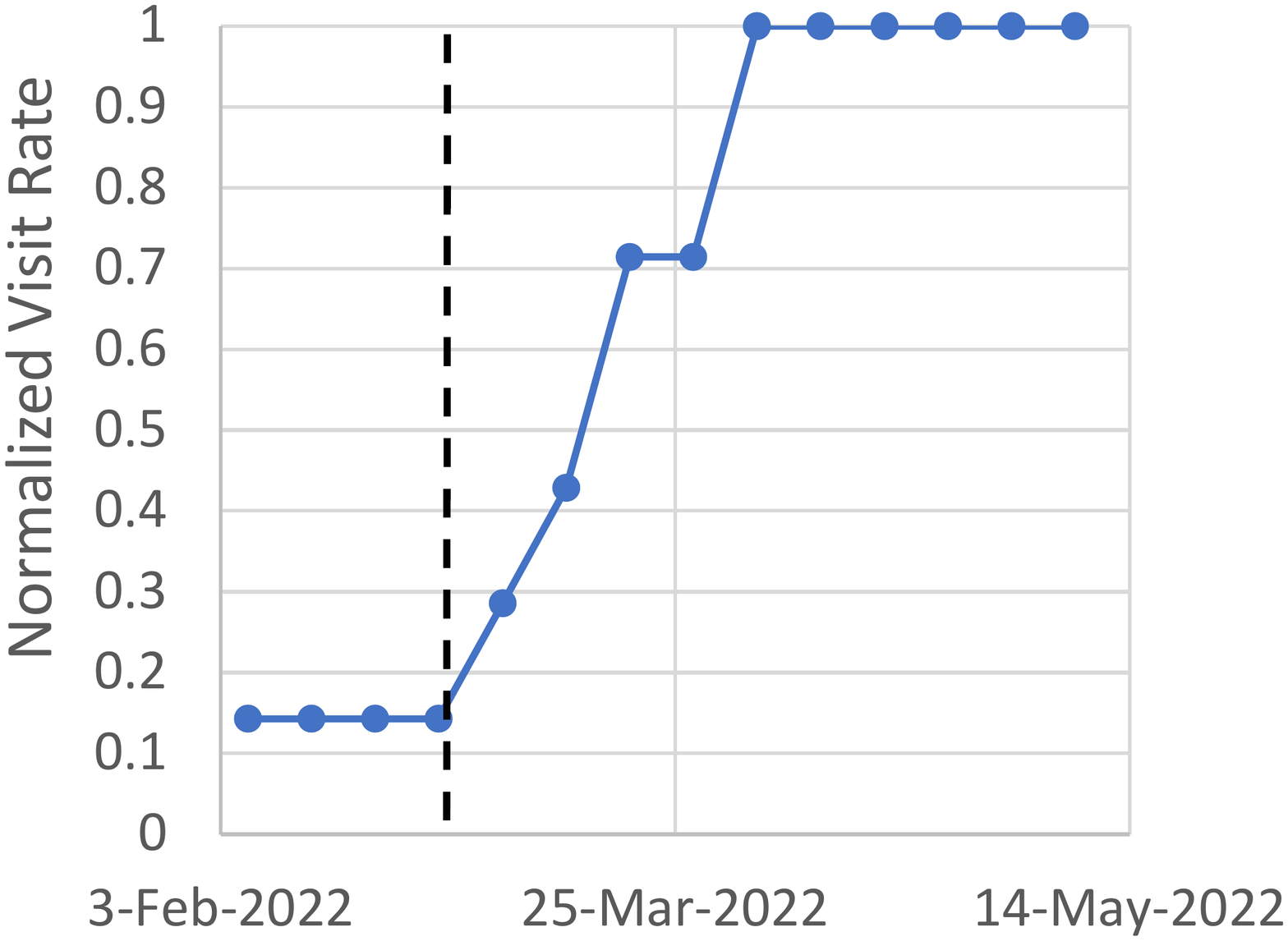}}
  \captionsetup{justification=centering}
  \caption{Germany}
  \label{fig:Goo1}
  \end{subfigure}%
  \begin{subfigure}[t]{.19\textwidth}
  \centering
  \fbox{\includegraphics[trim={5cm 2cm 4cm 2cm},clip,height=4\grafflecm]{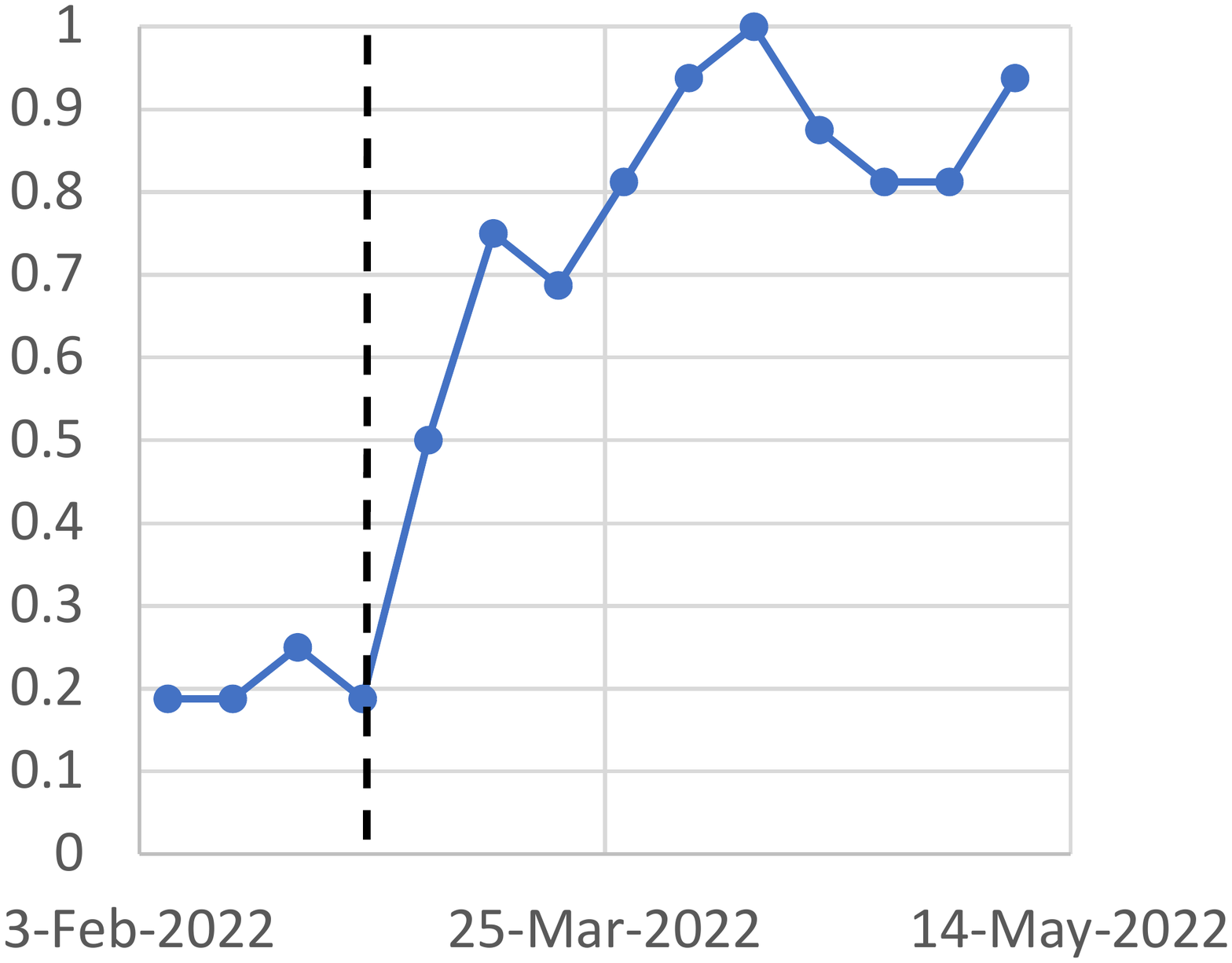}}
  \captionsetup{justification=centering}
  \caption{Poland}
  \label{fig:Goo2}
  \end{subfigure}%
  \begin{subfigure}[t]{.19\textwidth}
  \centering
  \fbox{\includegraphics[trim={5cm 2cm 4cm 2cm},clip,height=4\grafflecm]{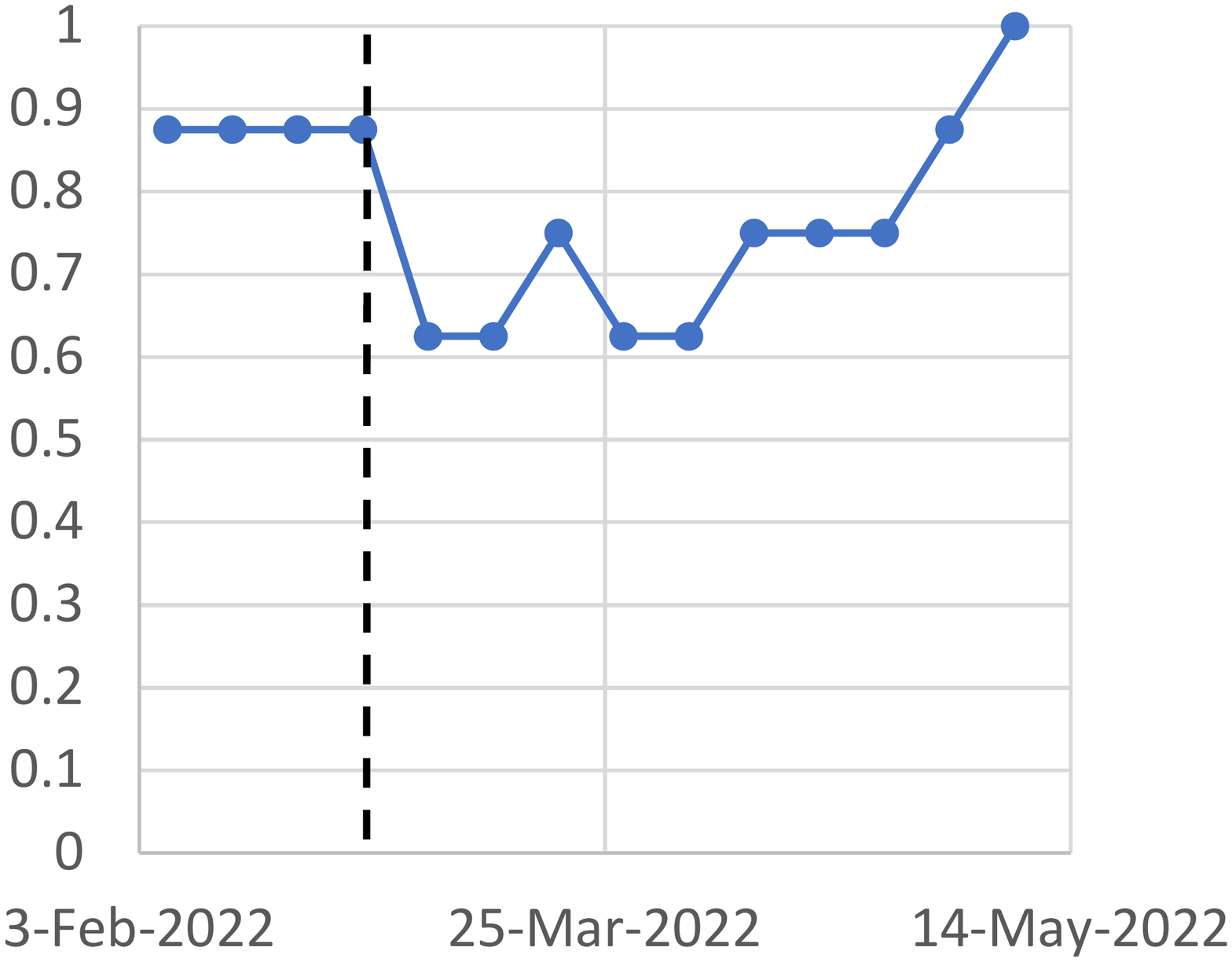}}
  \captionsetup{justification=centering}
  \caption{Russia}
  \label{fig:Goo3}
  \end{subfigure}%
  \begin{subfigure}[t]{.19\textwidth}
  \centering
  \fbox{\includegraphics[trim={5cm 2cm 4cm 2cm},clip,height=4\grafflecm]{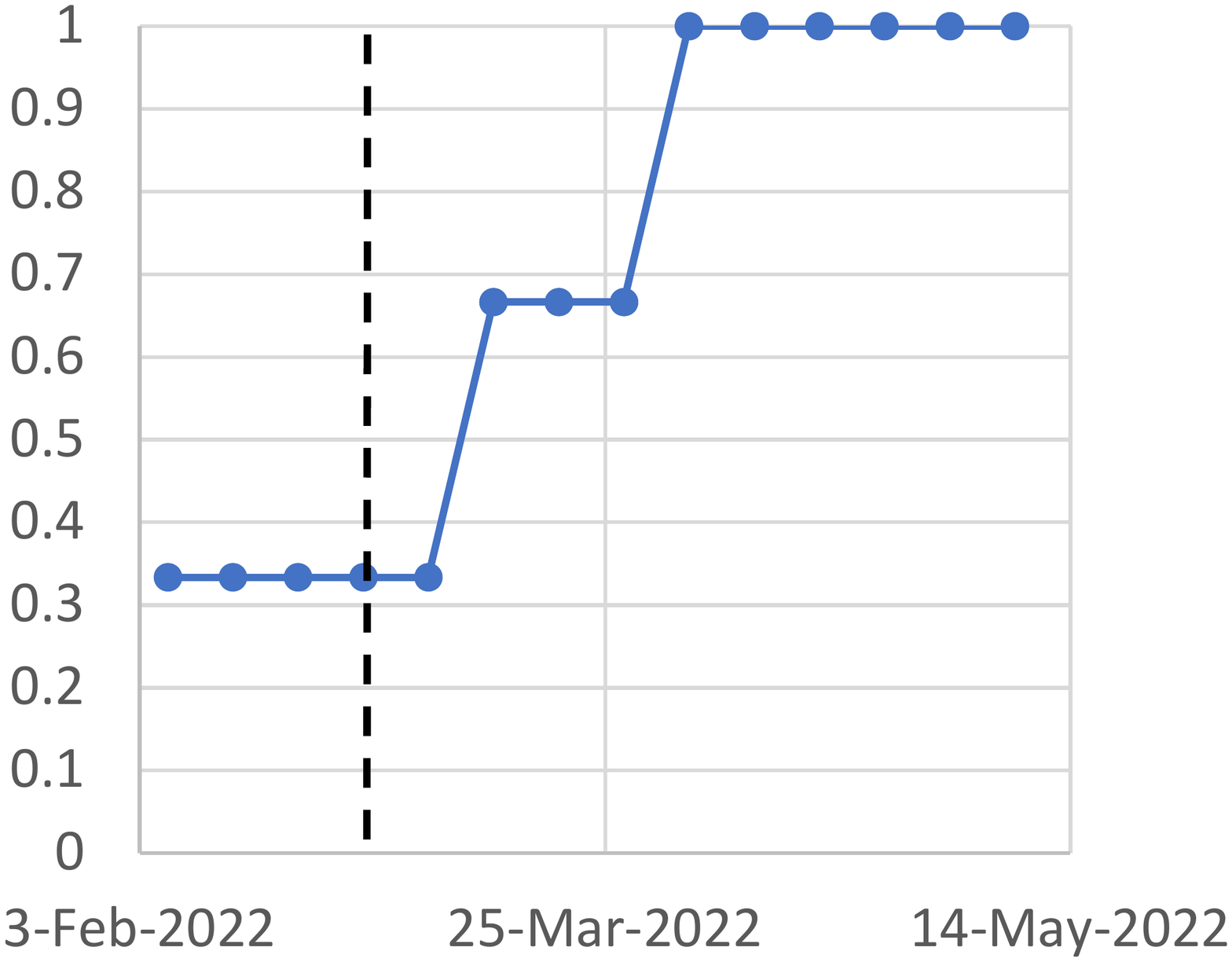}}
  \captionsetup{justification=centering}
  \caption{Netherlands}
  \label{fig:Goo4}
  \end{subfigure}%
  \begin{subfigure}[t]{.19\textwidth}
  \centering
  \fbox{\includegraphics[trim={5cm 2cm 4cm 2cm},clip,height=4\grafflecm]{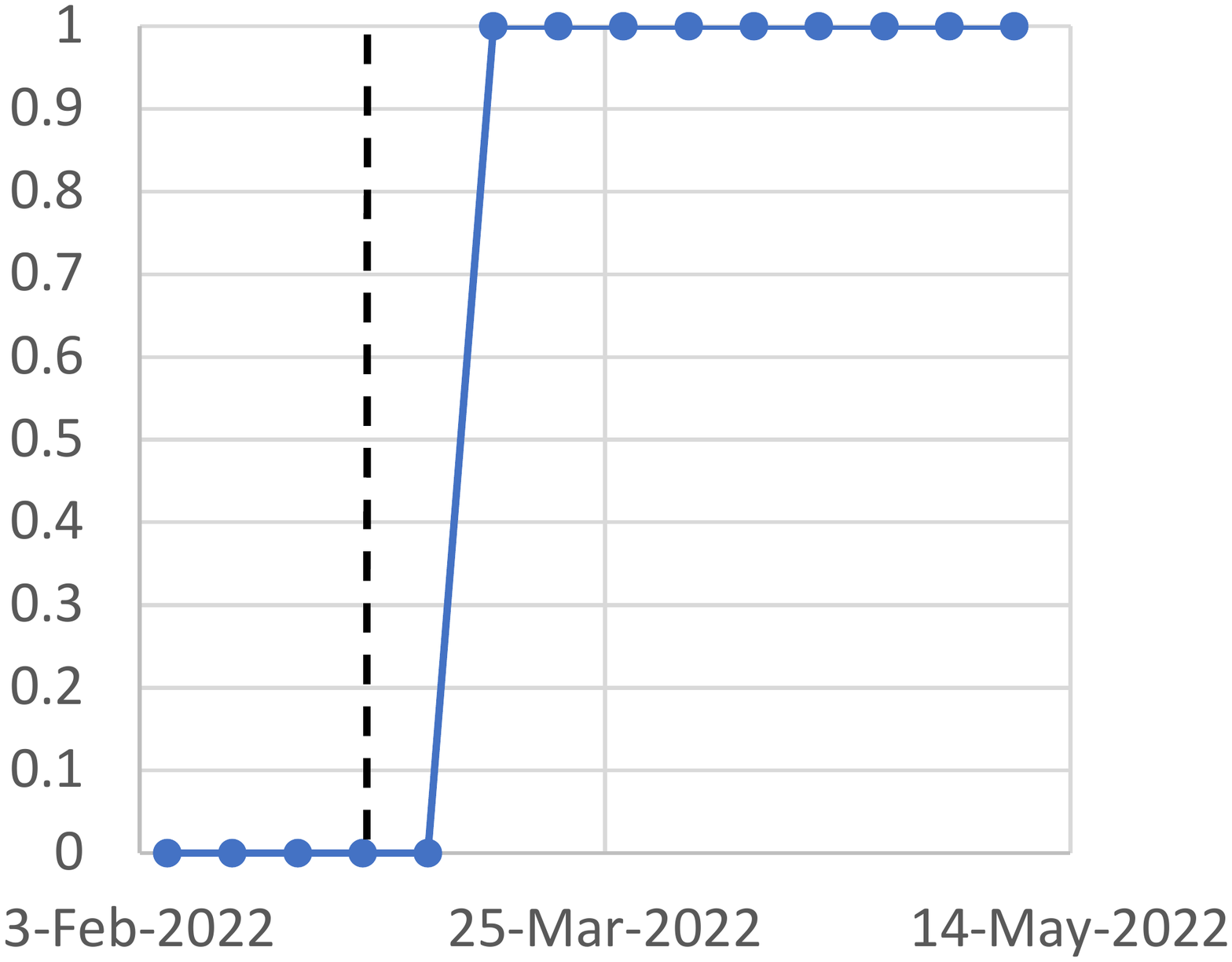}}
  \captionsetup{justification=centering}
  \caption{France}
  \label{fig:Goo5}
  \end{subfigure}%
	\captionsetup{justification=centering}
\ifdefined\CutSpace
  \caption{Normalized visit rate to google.com.ua from each country [Statcounter].}
\else
  \caption{Normalized visit rate to google.com.ua from each country [Statcounter]. \\ Dotted lines mark the beginning of the conflict.}
\fi
  \label{fig:CountryTrend}
\end{figure*}

\subsection{A Time Domain Perspective}
While Ukrainian website visit rates can provide a useful insight into Ukrainian presence in countries throughout the world, these figures are also affected by the Ukrainian diaspora prior to the war. According to~\cite{UNHCR} the number of Ukrainian citizens spread throughout the world in June 2020 was approximately 6.1 million. 

Notably, historic data about website visits can be used to assess the trends in the geographic location of Ukrainian website visits, and thus can assist in a time domain analysis of the trends in the flow of refugees.

\begin{sloppypar}
Our analysis focused on historic data that is publicly available; we found that Statcounter data includes \emph{search engine host market share} on a per-country basis, and specifically includes per-country data about the visits to google.com.ua. Notably, the data is expressed as the relative visit rate compared to local search engine requests. For example, 0.07\% of the search engine requests in Germany went to google.com.ua in May 2022. Conversely, the data on Figures~\ref{fig:SiteHist} and~\ref{fig:EstimatedHist} was expressed as the percentage from the total visits in the analyzed sites. Thus, in order to analyze the historic data, we normalize the data from each country, as shown in Figure~\ref{fig:CountryTrend}, illustrating the trends of the Ukrainian presence in each of these countries in a relative manner. For example, Figure~\ref{fig:Goo2} shows the vast influx of refugees into Poland on the first two weeks of the war, while the influx into Germany reached its peak on the third week of the war, as shown in Figure~\ref{fig:Goo3}.
\end{sloppypar}

Combining the two types of analytics provides a broader picture: website visit analytics (Figures~\ref{fig:SiteHist} and~\ref{fig:EstimatedHist}) provide the geographic distribution of Ukrainian presence, and historic data (Figure~\ref{fig:CountryTrend}) indicates the trends of the refugee flow.

\section{Related Work}
\label{RelWorkSec}
To the best of our knowledge, we are the first to publish academic research connecting Internet measurements to the Ukrainian refugee crisis. Several blogs that are related to the conflict in Ukraine have been published, among them \cite{UkrainianInternet} surveyed the status of the Internet infrastructure at the beginning of the conflict, and \cite{Resilience} studied the resilience of the Internet during the conflict.

The COVID-19 pandemic has widely affected the usage of the Internet over the last few years.
The studies published on this topic are an example of how Internet traffic might be used to analyze a worldwide crisis.  Specifically, \cite{candela2020impact} studied the impact of the COVID-19 pandemic on the Internet latency.  \cite{ukani2021locked} studied the changes of Internet traffic in on-campus dormitories during the pandemic lockdown, providing a focused lens on the behavior of undergraduate student population during the pandemic. In~\cite{FeldmannGLPPDWW20}, the authors examined the effect of government lockdowns on Internet traffic, finding a significant increase of 15-20\% of traffic volume.
\cite{bottger2020internet} provided a perspective of the scale of Internet traffic growth and how well the Internet coped with the increased demand as seen from Facebook’s edge network during the COVID-19 pandemic.  In \cite{lutu2020characterization}, the impact of the COVID-19 crisis on a UK mobile network operator is studied. \cite{feldmann2021year} characterized how Internet traffic and application demands change over a year in lockdown.

\section{Conclusion}
\label{ConcSec}
In this paper we analyzed how Internet measurement data can be used to map the Ukrainian refugee crisis. We demonstrated the unique footprint that the refugee crisis had on the Internet performance in Ukraine and around it, and introduced a novel approach to estimate the refugee distribution using publicly available information about website visits. Notably, the proposed methodology can achieve higher accuracy if more detailed data will become publicly available. Specifically, detailed historic data about website visit rates can provide a much deeper insight not only into the distribution of individuals throughout the world, but also about the trends and the flow of refugees. We hope that this work can contribute to the ongoing effort to support the refugees.

\ifdefined\TechReport
\else
\section*{Appendix: Ethics}
\label{EthicsSec}
This work does not raise any ethical issues. Specifically, any ethical issues that are related to the war in Ukraine are outside the scope of this work. From a privacy perspective, all the data that is analyzed in this work is publicly available, and the analysis produces statistics and trends, without compromising any privacy aspects.
\fi

\balance
\bibliographystyle{abbrv}
\bibliography{Uk}

\end{document}